\documentclass[aps,prd,showpacs,nofootinbib,twocolumn,superscriptaddress,floatfix]{revtex4}

\usepackage{amsmath,amssymb}
\usepackage[dvips, pdftex]{graphicx}
\usepackage[tight,footnotesize]{subfigure}
\usepackage[usenames]{color}
\usepackage[dvipsnames]{xcolor}
\usepackage{float}
\usepackage{wrapfig}
\usepackage{hyperref}
\usepackage{mathtools}

\DeclareGraphicsExtensions{.jpg,.mps,.pdf,.png,.eps} 

\renewcommand{\ij}{{}_{ij}} 
\renewcommand{\IJ}{{}^{ij}}

\newcommand{\gm}{\gamma}

\newcommand{\lp}{\left(}
\newcommand{\rp}{\right)}

\newcommand{\tgm}{{\tilde\gm}}

\newcommand{\tA}{{\tilde  A}}

\newcommand{\be}{\begin{equation}}
\newcommand{\ee}{\end{equation}}
\newcommand{\bea}{\begin{eqnarray}}
\newcommand{\eea}{\end{eqnarray}}

\newcommand{\rr}{\mathrm}

\newcommand{\mpl}{m_\rr{pl}}
\newcommand{\Mpl}{M_\rr{pl}}

\begin{document}
\title{
Inhomogeneous initial conditions for inflation: \\ A wibbly-wobbly timey-wimey path to salvation 
}

\author{Cristian Joana}
\email{cristian.joana@uclouvain.be}
\affiliation{Cosmology, Universe and Relativity at Louvain (CURL),
	Institut de Recherche en Mathematique et Physique (IRMP),
	University of Louvain,
	2 Chemin du Cyclotron,
	1348 Louvain-la-Neuve,
	Belgium}
\affiliation{Service de Physique Th\'eorique, Universit\'e Libre de Bruxelles (ULB), Boulevard du Triomphe, CP225, 1050 Brussels, Belgium.}

\author{S\'ebastien Clesse}
\email{sebastien.clesse@ulb.ac.be}

\affiliation{Service de Physique Th\'eorique, Universit\'e Libre de Bruxelles (ULB), Boulevard du Triomphe, CP225, 1050 Brussels, Belgium.}

\affiliation{Cosmology, Universe and Relativity at Louvain (CURL),
	Institut de Recherche en Mathematique et Physique (IRMP),
	University of Louvain,
	2 Chemin du Cyclotron,
	1348 Louvain-la-Neuve,
	Belgium}

	
\date{\today}

\begin{abstract}

We use the 3+1 formalism of numerical relativity to investigate the robustness of Starobinsky and Higgs inflation to inhomogeneous initial conditions, in the form of either field gradient or kinetic energy density.  Sub-Hubble and Hubble-sized fluctuations generically lead to inflation after an oscillatory phase between gradient and kinetic energies.  Hubble-sized inhomogeneities also produce contracting regions that may end up forming primordial black holes, subsequently diluted by inflation.  We analyse the dynamics of the preinflation era and the generation of vector and tensor fluctuations. Our analysis further supports the robustness of inflation to any size of inhomogeneity, in the field, velocity or equation of state.  
At large field values, the preinflation dynamics only marginally depends on the field potential and it is expected that such behaviour is universal and applies to any inflation potential of plateau-type, favoured by CMB observations after Planck.

\end{abstract}

\maketitle

\section{Introduction}\label{sec:intro}

In the inflationary paradigm, the Universe undergoes an early phase of nearly exponential, accelerated expansion.  Inflation naturally solves a series of problems of the standard cosmological model, including the flatness and horizon problems.   It is usually driven by one or several scalar fields that slowly roll along an almost flat direction of their potential.   Quantum fluctuations during inflation provide adiabatic and nearly scale-invariant curvature fluctuations, whose primordial power spectrum is today well constrained by cosmic microwave background (CMB) observations.   The latest results from Planck~\cite{Akrami:2018odb,Ade:2015lrj} favour single-field inflation with a plateau-like potential~\cite{Martin:2013nzq}, such as the Higgs/Starobinsky inflation model.  

Despite those great successes, the naturalness of the inflationary scenario has been questioned for about thirty years.  Indeed, inflation explains why the Universe is homogeneous over about $10^5$ Hubble volumes at the time of the last scattering.  But this would only push backwards the fine-tuning issue if the triggering of inflation requires homogeneous initial conditions over several Hubble volumes.  In such a case, the appealing and naturalness of inflation would be strongly reduced.

The question of how homogeneous must be the Universe prior to inflation has been addressed by several authors, with apparently contradictory results, so that the initial fine-tuning issue has been unclear until recently.   Linear density fluctuations certainly do not prevent the onset of inflation~\cite{Brandenberger:1990wu,Brandenberger:1990xu,Alho_2014, ALHO2011537}, but dealing with the fully relativistic non-linear dynamics of large inhomogeneities, including the backreactions on the Universe's expansion, is a much more complex problem.  This requires going beyond the linear theory of cosmological perturbations, for instance by using the gradient expansion formalism~\cite{Deruelle:1994pa, Azhar_2018} 
or non-perturbative approximations to capture some of the nonlinear backreactions \cite{Bloomfield_2019}.
In this context, methods of numerical relativity are well-suited~\cite{Goldwirth:1989vz,Laguna:1991zs,KurkiSuonio:1993fg} but their use has been limited for a long time by computational resources.   
Recently, numerical relativity in 3+1 dimensions has been used to study the early Universe cosmology in the context of inflation \cite{East:2015ggf,Clough:2016ymm,Aurrekoetxea_2020,PhysRevD.100.063543}, and possible alternatives \cite{PhysRevD.78.083537,Ijjas_2020}. In particular, the problem of initial conditions for inflation has been considered for several inflaton models and scalar field initial configurations~\cite{East:2015ggf,Clough:2016ymm,Aurrekoetxea_2020}.
Despite this progress, the required degree of homogeneity and the question of whether inflationary patches can emerge from a landscape of non-linear scalar field fluctuations have been only solved in some specific cases, and in general they are still controversial. First works obtained that inflation cannot start from sub-Hubble non-linear fluctuations, for instance by using numerical relativity in spherical symmetry \cite{Goldwirth:1989vz,Goldwirth:1990pm,Goldwirth:1989pr}, and  by using the gradient expansion formalism \cite{Deruelle:1994pa}.  An opposite result was obtained in Ref. \cite{KurkiSuonio:1993fg} by using the first numerical relativity simulations in 3+1 dimensions (see also Ref. \cite{Brandenberger:2016uzh}), which has been confirmed more recently in Ref. \cite{East:2015ggf}. 

In summary, despite recent progress, the mechanisms causing (or not) initial non-linear inhomogeneities to inflate are not fully understood yet, as well as their model dependence.
This work aims to contribute to paving the way to a better understanding of the fully relativistic non-linear preinflation dynamics, and thereby  a better view of viable and theoretically motivated inflationary models.  

In this paper, we investigate the inhomogeneous scalar field dynamics and the possible onset of inflation, with the use of numerical relativity in 3+1 dimensions.   For this purpose, we rely on the \texttt{GRCombo} code~\cite{Clough_2015}, based on the Baumgarte-Shapiro-Shibata-Nakamura (BSSN) formalism \cite{PhysRevD.52.5428,Baumgarte_1998,10.1143/PTPS.90.1}.  It is used to solve the full Einstein field equations together with the Klein-Gordon equation for a scalar field.
 
The BSSN formalism has been proved to be stable and efficient for a variety of problems, from the dynamics of black hole binaries to cosmological problems such as the gravitational collapse of cosmic strings \cite{Helfer_2019}, the non-linear collapse of matter inhomogenities~\cite{Rekier:2014rqa,Rekier:2015isa} and their backreactions on the Universe's expansion.

Our analysis focuses on the Higgs/Starobinsky inflation model.  This choice is well motivated for several reasons.  First, the potential has a single parameter, fixed through the CMB power spectrum normalization.   This restricts the parameter space to explore to the initial conditions of the field.  Second, it is the best favoured (and the simplest) inflation model after Planck~\cite{Martin:2013nzq}.   Third, the model has been considered in~\cite{Aurrekoetxea_2020}, which allows us to compare some of our results to the literature.

In particular, we reproduce the case of a Gaussian field fluctuation on top of a background field value lying in the slow-roll region.   But compared to previous work, our analysis has been extended to study more exhaustively the dynamics of the preinflation era and determine where inflation can take place and for which fluctuation sizes.  We consider not only the case of an inhomogeneous initial field, but also cases with inhomogeneous field velocity and equation of state, which have not been considered so far.  Some universal behaviours of the field and space-time dynamics are identified, depending on the characteristic fluctuation sizes.  Our results further support the robustness of inflation against various configurations of initial conditions, and for sub-Hubble and Hubble-sized fluctuations they better emphasize a universal behaviour in the form of an oscillating equation of state, a signature of the respectively quick or slow wobbling between field gradient and kinetic terms that alternatively dominate the total energy density.

For each case considered, we monitor the evolution of all geometrical quantities, the scalar field and its velocity.  We identify the conditions under which inflation can be triggered in some parts of the lattice, whereas other parts can undergo a gravitational collapse leading to black hole formation.   Finally, we clarify and explain why previous works have led ostensibly to different conclusions, which is related to the time at which the initial Hubble scale is defined.  We emphasize that the level of initial inhomogeneity is naturally restricted if the energy density is initially dominated by field gradients.   Inflation is generally the natural outcome, except in regions where the field Laplacian is maximal that can start contracting and collapse into preinflation black holes.

The paper is organized as follows:  In Section~\ref{sec:prevwork} we review previous results in the topic of the inhomogeneous initial conditions of inflation.  The Higgs and Starobinsky models are introduced in Section~\ref{sec:potential}. The BSSN formalism of numerical relativity is detailed in Section~\ref{sec:BSNN} and
its link to coarsed-grained cosmology and metric perturbations is explained in Section~\ref{sec:cors_universe}. 
Section~\ref{sec:inital_conditions} describes the considered initial conditions.  Our results are presented in Section~\ref{sec:results} and their implications are discussed in Section~\ref{sec:discussion}.  Our conclusion and the perspectives of this work are presented in {Section}~\ref{sec:ccl}.  
In appendix, we provide more technical details on the convergence tests to check the stability and validity of our simulations.

\section{Summary of previous work}  \label{sec:prevwork}

There exist only a few references having investigated the initial inhomogeneity problem of inflation.   In this section, we give a brief and general overview of previous works on this topic.
The question of how generic or fine-tuned are \textit{homogeneous} initial conditions leading to inflation is another related issue, also controversial, and we will make some connections to this problem -- in particular to the slow-roll attractor solution and the dynamics of the preinflation phase in the presence of a large kinetic term -- for plateau inflation.  We refer the interested reader to the recent literature, see e.g.~\cite{Linde:2017pwt,Finn:2018krt,Chowdhury:2019otk,Tenkanen:2020cvw}.

The first attempts to study the problem of \textit{inhomogeneous} initial conditions using numerical relativity are due to Goldwirth and Piran in 1989 and 1990 \cite{Goldwirth:1989vz,Goldwirth:1989pr,Goldwirth:1991rj}. 
The numerical method was introduced in Ref. \cite{Goldwirth:1989vz} and their results were presented in Refs. \cite{Goldwirth:1989pr,Goldwirth:1991rj} for five scalar field potentials:  large-field inflation with quadratic and quartic potentials, and small-field inflation with a quartic or a Coleman Weinberg (CW) potential. \footnote{In Refs. \cite{Goldwirth:1989pr,Goldwirth:1991rj}, large-field and small-field inflation were referred to using the older terms \textit{chaotic} and \textit{new} inflation, respectively.}   
They focused on spherically symmetric cosmologies and found that only sufficiently large inhomogeneities, at least of the size of the Hubble radius, can lead to large-field inflation.  They gave an estimate of the fluctuation size in units of $H^{-1}$ preventing the onset of inflation.  For all these models, the onset of inflation required homogeneity over several horizon sizes, with the notable exception of the CW potential.  However, for small-field inflation, the onset of inflation requires tiny values of the mean scalar field and its fluctuations.  Thus, inflation seems to be less natural in small-field inflation than in large-field inflation.  Sub-Hubble fluctuations typically did not lead to inflation, because the slow-roll regime cannot be reached before the mean-field value reaches the bottom of the potential. They nevertheless pointed out that for sub-Hubble fluctuations, inflation could be triggered if scalar field oscillations are damped down to a sufficiently large homogeneous field value.  But, due to obvious computational limitations, they were not able to cover full ranges of fluctuation sizes and amplitudes.  They also did not cover the interesting parameter range for the considered models since CMB observations were not available at the time.  Furthermore, spherical symmetry does not capture all the possible general relativistic effects, and so the application of their results to more general fluctuations can be questioned.  Finally, they did not consider cases where the initial density is strongly dominated by kinetic or gradient terms.  As a consequence, the analytical approximation proposed by Goldwirth in~\cite{Goldwirth:1990pm}, stating that the comoving size of inhomogeneities $\Delta$ needs to be such that
\be
a \Delta > \sqrt{ \frac{ 8 \pi}{3 } } \frac{\delta \varphi}{H_{\rr{inf}} \mpl} 
\ee
where $a$ is the initial scale factor, assumes that the energy density is dominated by the potential and the field gradients are suppressed.  If this is true at the onset of inflation, this assumption can be violated in the preinflation era, with $H_{\rr{inf}} \ll H_{\rm ini} \lesssim m_{\rm Pl} $.  

An impressive study of the problem, using 3+1 numerical relativity, has been achieved in 1993 by Kurki-Suonio \textit{et al.}   As in previous works, they used the York's procedure \cite{York:1971hw,York:1972sj} to solve the initial condition problem, for a large-field quartic potential.   For inhomogeneous runs, they correctly related the Hubble horizon size to the total density, including the field gradients and velocities.  For their initial conditions, the density is actually dominated by kinetic terms, and $H\sim m_{\rm pl}$ initially.  They ran a few 3D simulations, both for super-Hubble and slightly sub-Hubble fluctuations.   In the latter case they first observed field oscillations, and when the fluctuations become super-Hubble expansion occurs in a homogeneous way and inflation can take place before the field reaches the bottom of the potential.  They concluded that inflation can arise from non-linear field fluctuations of the order of the Hubble horizon and beyond.  However, they restricted their analysis to a potential shape -- and set of parameters -- that is now strongly disfavoured by CMB observations and only considered initial conditions with sub-dominant field gradients compared to the kinetic and potential terms.  Nevertheless,  their analysis certainly remains very impressive given that it was performed in 3D, with limited computational resources compared to those currently available.

As an alternative to full numerical relativity methods, Deruelle and Goldwirth used in 1995 a long wavelength iteration scheme ~\cite{Deruelle:1994pa}, also referred to as the gradient expansion formalism, in order to determine how large the initial homogeneities of a massive scalar field can be without preventing inflation from setting in.  They found that homogeneity over patches of the order of, or larger than the local Hubble radius, is a general condition needed for inflation, but such a method does not allow to study the evolution of strong initial inhomogeneities.  It nevertheless provides an understanding of the factors controlling the system behavior, without assuming any spatial symmetry.  

After a gap of about twenty years, the problem of inhomogeneous initial conditions for inflation has seen a renewed interest in the recent literature
\cite{Ijjas_2013,Guth_2014,Ijjas_2016}\cite{Chowdhury:2019otk}. In parallel, inhomogeneous initial field values in multi-field inflation models were considered in Ref. \cite{Easther:2014zga}, but without including gravitational backreactions.  Their analysis particularly focused on hybrid models of inflation, extending the work of Refs. \cite{Clesse:2009ur,Clesse:2008pf}.  

In the fall of 2015, East \textit{et al.} released a new analysis of the initial homogeneity problem~\cite{East:2015ggf} based on numerical relativity.  They considered three scalar field potentials:  a constant, a smooth step. and a notch potential, with the latter describing a family of cosmological attractors.  They studied initial conditions dominated by the field gradient energy and used 3+1 lattice simulations in full general relativity, thus extending the initial work of Kurki-Suonio \textit{et al.}  They found that field fluctuations initially smaller than the Hubble radius but contained in a flat region of the potential can lead to inflation, after the gradient and kinetic field energy is diluted by expansion.  They also found that, at the same time underdense regions lead to inflation, overdense regions can collapse and form preinflation black holes (PIBHs).  

In Ref. \cite{Clough:2016ymm}, Clough \textit{et al.} used numerical relativity in 3+1 dimensions to study the robustness of initial conditions leading to inflation, for different inflationary models.  In particular, they compared large-field to small-field scenarios.  Their results suggest that it is much less natural to get inflation in small-field models even when the gradient energy is initially subdominant.   This result is nevertheless mitigated by the fact that initial field values outside the slow-roll regions of the potential can lead to inflation.  For large-field models with relatively flat initial hypersurfaces, they also confirmed that PIBH formation does not prevent the onset of inflation in other regions.  In a following paper~\cite{Clough_2018}, the authors considered inhomogeneities in the metric sector with tensor modes, while keeping the scalar field rather homogeneous. They noticed a reduction in the duration of inflation for small-field models, however suppressed for large tensor modes due to a large Hubble fiction. Large-field models are not strongly affected by the tensor perturbations. Gravitational collapse due to tensor modes was also reported.

Most recently, the effect of the potential shape on initial scalar field gradients was further explored in~\cite{Aurrekoetxea_2020}.  Convex potentials were found to be more robust than concave ones for sub-Planckian characteristic scales, for which the field can be dragged-down to the bottom of the potential with a significant loss of efolds.  Super-Planckian scales can more generally lead to long enough inflation, also for concave potentials. They suggest that the onset (or not) of sufficient inflation can be inferred from an analytical criterion, consisting in finding the critical scalar field amplitude for which the dragging-down of the potential overcomes the pull-back effect of the gradient pressure.  

Some of these considerations were summarized early 2016 by Brandenberger in a short and general review on the issue of the initial conditions for inflation~\cite{Brandenberger:2016uzh}, where they reported on the possible solutions to an initial fine-tuning problem, and emphasised that slow-roll appears to be a local attractor for large-field models, in contrast to small-field models.

In summary, the current status of the paradigm is that in large-field and plateau potentials, non-linear initial field Hubble-sized fluctuations do not prevent the onset of inflation.  The preinflation dynamics of sub-Hubble and Hubble-sized fluctuations is nevertheless not yet fully understood, in particular in the case of PIBH formation.  It is also uncertain whether these conclusions still apply for initially inhomogeneous field velocity or any other type of initial conditions.  The dynamics of the Higgs/Starobinsky model that is the simplest and one of the best favoured model nowadays, were not explored in detail until now.

\section{Higgs/Starobinsky inflation  \label{sec:potential}}

The Higgs inflation model~\cite{Bezrukov:2007ep} identifies the inflaton as the Standard Model Brout-Englert-Higgs field, but it is non-minimally coupled to gravity in order to provide a sufficiently flat potential to realise inflation.   It is the simplest inflationary model which relates to the Standard Model of particle physics. The Lagrangian includes an extra term $\xi H^\dagger H R$, where $H$ is the Higgs field, $R$ is the Ricci scalar, and $\xi$ is the only parameter of the model.   This term is generated automatically by quantum corrections in curved space-time.  In the Einstein frame, the action is the one of a minimally coupled scalar field with the following potential,
\be
V(\varphi) = \Lambda^4 \left( 1 - {\rm e}^{-\sqrt{2/3}  \varphi  / \Mpl} \right)^2~,
\ee
where $\Mpl$ is the reduced Planck mass, $\Lambda^4 \equiv \Mpl^4 \lambda / (4 \xi^2)$ becomes the unique potential parameter, $\lambda$ being a constant characterizing the model.  The CMB observations allow to fix $\Lambda \simeq 3.1 \times 10^{-3} \Mpl $~\cite{Martin:2013nzq}.  
At large field values, the potential has a plateau allowing for slow-roll inflation.   Slow-roll conditions are violated, and inflation ends at $\varphi_{\rm end} \approx 0.94 \Mpl$. At bout $\Delta N_* \simeq 62 $ efolds before the end of inflation, observable scales leave the Hubble radius, at a field value $\varphi_* \approx 5.48\ \Mpl$.

The Starobinsky model of inflation is a scalar-tensor theory with $f(R) = R + \epsilon R^2 / \Mpl^2 $, which in the Einstein frame has the same linear dynamics and effective field potential as Higgs inflation.  Therefore, even if the non-linear dynamics of scalar field and BSSN variables during the preinflation era can differ between the Jordan and Einstein frame for the Starobinsky model, showing that inflation is reached in the Einstein frame is a sufficient condition to guarantee that inflation is also reached in the Jordan frame.
Although these models have different reheating mechanisms, and eventually distinguishable observable predictions, we are not interested in this issue in this paper.  Finally, let us point out that the Higgs/Starobinsky model is the
best favored slow-roll inflation model after Planck~\cite{Martin:2013nzq}.

\section{BSSN formalism of numerical relativity}\label{sec:BSNN}

In this work, we solve the BSSN formulation of the Einstein equations using \texttt{GRChombo}~\cite{Clough_2015}, a multipurpose numerical relativity code. In the context of the 3+1 decomposition of General Relativity, the line element can be written as
\be\label{timeline}
\rr d s^2  = - \alpha^2 \rr d t^2 + \gamma_{ij}(\rr d x^i + \beta^i \rr d t)(\rr d x^j + \beta^j \rr d t)
\ee
where $\gamma\ij$ is the metric of the 3-dimensional hypersurface, and the   lapse and shift gauge parameters are given  by $\alpha(t)$ and $\beta^i(t)$ {respectively}.  A further conformal decomposition of the 3-metric follows,
\be
\gm\ij = \frac1\chi \tgm\ij = \psi^4\tgm\ij \quad \text{with } \text{ det}(\tgm\ij) =  1 ~,  
\ee
where $\chi$ and $\psi$ are two different parametrisations of the metric conformal factor. While the former is used during the temporal integration, the latter is preferred when constructing the initial conditions. The extrinsic curvature is thus split in $\tA\ij$ and $K$, respectively, the conformal traceless part and its trace,  
\be
K\ij = \frac1\chi \left( \tA\ij +\frac13\tgm\ij K\right)~.
\ee
In addition, the first spatial derivatives of the metric are considered as dynamical variables
\be
\tilde\Gamma^i \equiv \tgm^{jk} \tilde\Gamma^i_{jk} = - \partial_j\tgm\ij ~,
\ee
where $ \tilde\Gamma^i_{jk} $ are the Chritoffel symbols associated with the conformal metric $ \tilde\gamma_{ij} $.

\subsection{Evolution equations}

The evolution equations for the BSSN variables are then given by 
\begin{align} 
&\partial_t\chi=\frac{2}{3}\,\alpha\,\chi\, K - \frac{2}{3}\,\chi \,\partial_k \beta^k + \beta^k\,\partial_k \chi ~ , \label{eqn:dtchi2} \\
&\partial_t\tilde\gamma_{ij} =-2\,\alpha\, \tA_{ij}+\tilde \gamma_{ik}\,\partial_j\beta^k+\tilde \gamma_{jk}\,\partial_i\beta^k \nonumber \\
&\hspace{1.3cm} -\frac{2}{3}\,\tilde \gamma_{ij}\,\partial_k\beta^k +\beta^k\,\partial_k \tilde \gamma_{ij} ~ , \label{eqn:dttgamma2} \\
&\partial_t K = -\gamma^{ij}D_i D_j \alpha + \alpha\left(\tilde{A}_{ij} \tilde{A}^{ij} + \frac{1}{3} K^2 \right) \nonumber \\
&\hspace{1.3cm} + \beta^i\partial_iK + 4\pi\,\alpha(\rho + S) \label{eqn:dtK2} ~ , 
 \end{align}
 \begin{align} 
&\partial_t\tA_{ij} = \left[- \chi D_iD_j \alpha + \chi \alpha\left( R_{ij} - 8\pi\, \,S_{ij}\right)\right]^\textrm{TF} \nonumber \\
&\hspace{1.3cm} + \alpha (K \tA_{ij} - 2 \tA_{il}\,\tA^l{}_j)  \nonumber \\
&\hspace{1.3cm} + \tA_{ik}\, \partial_j\beta^k + \tA_{jk}\,\partial_i\beta^k \nonumber \\
&\hspace{1.3cm} -\frac{2}{3}\,\tA_{ij}\,\partial_k\beta^k+\beta^k\,\partial_k \tA_{ij}\,   \label{eqn:dtAij2} ~, \\ 
&\partial_t \tilde \Gamma^i=2\,\alpha\left(\tilde\Gamma^i_{jk}\,\tA^{jk}-\frac{2}{3}\,\tilde\gamma^{ij}\partial_j K - \frac{3}{2}\,\tA^{ij}\frac{\partial_j \chi}{\chi}\right) \nonumber \\
&\hspace{1.3cm} -2\,\tA^{ij}\,\partial_j \alpha +\beta^k\partial_k \tilde\Gamma^{i} \nonumber \\
&\hspace{1.3cm} +\tilde\gamma^{jk}\partial_j\partial_k \beta^i +\frac{1}{3}\,\tilde\gamma^{ij}\partial_j \partial_k\beta^k \nonumber \\
&\hspace{1.3cm} + \frac{2}{3}\,\tilde\Gamma^i\,\partial_k \beta^k -\tilde\Gamma^k\partial_k \beta^i - 16\pi\,\alpha\,\tilde\gamma^{ij}\,S_j ~ , \label{eqn:dtgamma2}
\end{align} 
where the superscript $\rm{TF}$ denotes the trace-free parts of tensors.  The 3+1 decomposition of the energy-momentum tensor $T^{\mu\nu}$ gives
\bea \label{3+1sources}
    \rho  &=& n_\mu n_\nu T^{\mu\nu} ~,\\
    S_i  &=& -\gamma_{i\mu} n_\nu T^{\mu\nu} ~,\\ 
    S_{ij}  &=&  \gamma_{i\mu} \gamma_{j\nu} T^{\mu\nu} ~,\\ 
    S &=& \gamma\IJ S\ij ~,
\eea
where $n_\mu=(-\alpha,\vec 0)$ is the unit normal vector to the three-dimensional slices.

The Hamiltonian and Momentum constraints, 
\begin{align}
\mathcal{H} & = R + K^2-K_{ij}K^{ij}-16\pi \rho = 0\, , \label{eqn:Ham}  \\
\mathcal{M}_i & = D^j (K_{ij} - \gamma_{ij} K) - 8\pi S_i =0\, .  \label{eqn:Mom}
\end{align}
are only solved explicitly when constructing initial data.  They are also monitored during the time evolution in order to ensure that there is no  significant deviations from General Relativity.

\subsection{Scalar field equations}
For a single scalar field $\varphi$,  the energy-momentum tensor is given by
\begin{equation}
T_{\mu\nu} = \partial_\mu \varphi\, \partial_\nu \varphi - \frac{1}{2} g_{\mu\nu}\, \partial_\lambda \varphi \, \partial^\lambda \varphi - g_{\mu\nu} V(\varphi) \,
\end{equation}
where $V(\varphi)$ is the scalar field potential. The scalar field dynamics is governed by the  the Klein-Gordon equation, split into two first order equations for the field and its momentum $\Pi_{\rm M}$
\begin{align}
\partial_t \varphi &= \alpha \Pi_{\rm M} +\beta^i\partial_i \varphi \label{eq:dtvarphi} ~ , \\
\partial_t \Pi_{\rm M} &= \beta^i\partial_i \Pi_{\rm M} + \alpha\partial_i\partial^i \varphi + \partial_i \varphi \, \partial^i \alpha \\
& \ +\alpha \left( K\Pi_{\rm M}-\gamma^{ij}\Gamma^k_{ij}\partial_k \varphi - V'(\varphi) \right) \label{eq:dtPi} ~ ,
\end{align} 
where the superscript ($'$) denotes the derivative with respect to the field.

\subsection{Gauge choice and singularity avoidance }%

The gauge parameters are initially set to $\alpha=1$ and $\beta^i=0$ and then evolved in accordance with the \textit{moving puncture gauge} \cite{Baker_2006, Campanelli_2006}, for which evolution equations are
\begin{eqnarray}
\partial_t \alpha &=& -\eta_\alpha \alpha K +  \,\beta^i\partial_i \alpha \ , \label{eqn:alphadriver}\\
\partial_t \beta^i &=& B^i\, \label{eqn:betadriver},\\
\partial_t B^i &=& \frac34\, \partial_t \tilde\Gamma^i - \eta_B\, B^i\ \,, \label{eqn:gammadriver}
\end{eqnarray}
where the constants $\eta_\alpha$ and $\eta_B$ are conveniently chosen to improve the numerical stability.  This way, $\alpha$ and $\beta^i$ are boosted in the problematic regions with strongly growing extrinsic curvature and spatial derivatives of the three-metric $\tilde \gamma_{ij}$.  
The goal of this gauge is to prevent the code from resolving the central singularity of any black hole that may eventually form.

\section{Link to coarse-grained Cosmology  \label{sec:cors_universe} }

One can map the BSSN variables into more typical cosmological variables (scale factor $a$, Hubble rate $H$ and equation of state $w$) in the separate universe assumption, corresponding to homogeneity or the super-Hubble limit of field and metric fluctuations.  If one assumes $\beta^i = 0$ but keeps $\alpha$ as an arbitrary gauge choice,  the scale factor can be defined as  $a^2 = \chi^{-1}$ and, using Eq. (\ref{eqn:dtchi2}), the inhomogeneous analogue of the first Friedmann equation reads
\be
H \equiv \frac{\dot a}{a} = -\frac{1}3 \alpha K ~, \label{eq:HubbleParam}
\ee
where a dot denotes the derivative with respect to cosmic time.  By taking its time derivative and using Eq.~(\ref{eqn:dtK2}), one gets the equation for the acceleration of the expansion of Universe,
\begin{align}
\frac{\ddot a}{a} &= -\frac\alpha3\left( \frac{\dot\alpha}\alpha K - D^iD_i\alpha +  \alpha \left(\mathbf{\aleph} + 4\pi T \right) \right) ~,\label{eq:adotdot} \\[2mm]
\mathbf{\aleph} &\equiv  \tA\ij \tA\IJ ~,\\
T &\equiv \rho+S =  3\rho \left(\frac13 + \omega_\varphi\right) ~.
\end{align}
The first two terms are gauge-related terms that vanish in the \textit{synchronous} gauge (where $\alpha = 1$). The $\mathbf{\aleph}$ is a new geometrical term that vanishes in the homogeneous and isotropic case, and $T$ is the trace of the energy-momentum tensor.
The latter can be written in terms of the scalar field equation of state  $\omega_\varphi (t)$, whose value depends on the dominant term in the scalar field's energy density
\begin{align}
    \rho &\equiv  \frac12 \Pi_{\rm M}^2 + \frac12 \partial_i \varphi\, \partial^i\varphi  + V(\varphi) \\
    &=\rho_{\rm kin}  + \rho_{\rm grad}  + \rho_{\rm V} ~,
\end{align}
corresponding to the kinetic, gradient and potential energy density, respectively. Thus, one can identify three limiting cases:
\be
 \omega_\varphi (t) \simeq \left\{
 \begin{array}{rl}
    1   &  \quad\rightarrow \quad \rho \simeq \rho_{\rm kin}  ~, 
    \\
    -1/3  & \quad\rightarrow \quad \rho \simeq \rho_{\rm grad} ~,
    \\
    -1  & \quad\rightarrow \quad \rho \simeq \rho_{\rm V}  ~.
 \end{array} \right.
\
\ee

By interpreting Eq. (\ref{eq:adotdot}), and neglecting the gauge effects, (i.e.  $\dot \alpha =0$, and assuming $D^iD_i\alpha \approx 0$),  we can infer that the conditions for allowing a positive accelerated expansion of the Universe, ($\ddot a > 0$),  are
\begin{gather}
\omega_\varphi  < -\frac13 ~, \label{eq:dec.cond1}
\\
\mathbb{\aleph} <  \left| 12\pi\rho \left( \frac13 + \omega_\varphi\right) \right|  ~.
\label{eq:dec.cond2}
\end{gather}

As a result, one finds that the expansion rate of the Universe is governed by the interplay of the energy density $\rho$, the equation of state $\omega_\varphi$ and a geometrical parameter $\mathbb{\aleph}$. We remark that, even in scenarios where $\rho > \mathbb{\aleph} > 0$, the  $\mathbb{\aleph}$-term will become dominant when $\omega_\varphi \approx -\frac13$.  
\\ 

\begin{figure}[b]
\begin{center}
\includegraphics[width=8.5cm]{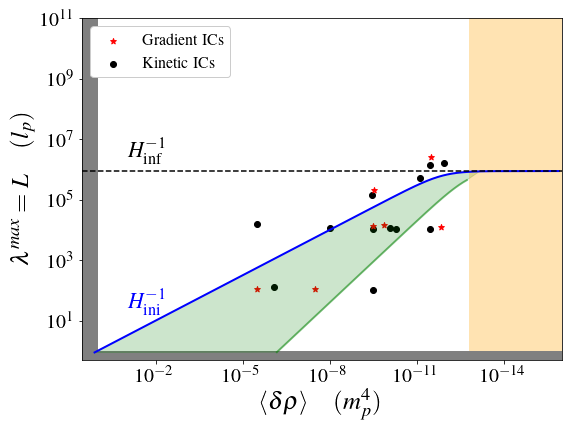}
\caption{ Representation of the considered initial conditions in terms of the maximal physical mode (delimited by the lattice size $L$) and the characteristic mean value of the inhomogeneous energy density. Black and red dots represent one simulation with respectively gradient or kinetic dominated initial conditions. The size of the Hubble scale at origin is represented by the continuous blue line, while the horizontal dotted line indicates the Hubble length at the onset of inflation at field values on the potential plateau. 
The green shaded area indicates the sub-Hubble region where dominant extrinsic curvature modes are produced at later simulated times.
The orange shaded region indicates the inflationary domain. 
The black shaded areas delimit the super-Planckian energy scales, which are excluded in our simulations.\label{fig:scale_diagram}}
\end{center}
\end{figure}

We can interpret $\mathbb{\aleph}$ as an energy density associated to perturbations in the spatial metric.  One sees that Eq.~(\ref{eqn:dttgamma2}) reduces to 
\be
\partial_t \tilde \gamma_{ij} = -2 \alpha \tilde A_{ij}
\ee
and therefore,
\be
 \mathbb{\aleph} \propto \dot\tgm\ij \dot\tgm\IJ
\ee
which is constituted by vector and tensor modes. The $\dot\tgm\ij$ is not related to scalar perturbations because $\tilde A_{ij}$ is traceless by definition.  Within a flat background, one would interpret them as a cosmic shear and gravitational waves.
In the absence of source terms in the evolution equation of $\tilde A_{ij}$ (i.e. $R\ij^{TF}, S\ij^{TF} \approx 0$, in Eq.~\ref{eqn:dtAij2}) they are quickly decaying as the universe expands, where the cosmic shear goes like $\mathbb{\aleph}_{\rm shear} \propto a^{-6}$, and gravitational waves like $\mathbb{\aleph}_{\rm GW} \propto a^{-4}$. 
However, such definitions are ill-defined in highly non-linear systems as the source terms of $\mathbb{\aleph}$ are no longer negligible and ultimately $\mathbb{\aleph}$ cannot be defined as a gauge-invariant quantity.  So we already point out that the importance and evolution of $\mathbb{\aleph}$ can only be revealed in fully relativistic $3+1$ simulations. 
\\

Because $\mathbb{\aleph}$ is strictly defined by the (traceless) extrinsic curvature, in this paper we also refer to its sub-Hubble modes as extrinsic curvature modes (ECMs).

\section{Initial conditions  \label{sec:inital_conditions}}

We perform our simulations of the preinflationary era for various sets of initial conditions (ICs).  
In all cases, we follow most of previous works and assume initial conformal flatness, i.e. $ \tgm\ij(t_0) = \delta\ij$ and $\tilde\Gamma^i(t_0) = 0$. Under this assumption, and by setting $\tA\ij(t_0) = 0$, the constraint equations (\ref{eqn:Ham})-(\ref{eqn:Mom}) are greatly simplified and can be rewritten as 
\begin{align}
 {\cal H}^* &= - \nabla^2 \psi + \psi^5 \left( \frac23 K^2 - 16\pi\rho  \right) ~, \label{eq:HamBBSN} \\
\tilde {\cal M}_i^* &= \frac23 \partial_i K + 8\pi S_i   \label{eq:MomBBSN} \\
&\text{with } ~ S_i = -\Pi_{\rm M}\partial_i \varphi 
~. \nonumber
\end{align}

Defining the perturbed energy density as $\delta \rho \equiv \rho - \rho_V$,  where initially $\delta \rho \gg \rho_V$, we  
then consider the following types of ICs:  
\begin{enumerate}
    \item When the energy density is dominated by field gradients, i.e.  $\delta \rho (t_0) = \rho_{\rm grad}$, 
    \item When the energy density is dominated by inhomogeneous field velocities,  i.e. $\delta\rho (t_0) = \rho_{\rm kin}$ 
    \begin{enumerate}
    \item with sub-dominant mean value,  $  \langle \Pi_{\rm M} \rangle  \sim 0  $ 
    \item with dominant mean value, $ | \langle \Pi_{\rm M} \rangle | \simeq   \Pi_{\rm M}  $.
    \end{enumerate}
\end{enumerate}
With these choices, the momentum density initially vanishes, $S_i =0$,  and therefore the momentum constrain is trivially satisfied if one considers an  homogeneous $K$.  Here, we have use the $\langle ... \rangle$ brackets to represent the mean value of a given function (i.e, $\theta$), averaged across the physical volume ${\mathcal V}$ represented {by} the whole lattice.  {For instance,}
\be \label{eqn:Kini}
\langle \theta \rangle \equiv \frac{1}{{\cal V}} \int \theta \, \rr d {\cal V}~,
\ee
{which takes into account the inhomogeneous physical volume of lattice cells that depends on the local value of the conformal factor at a given time.}  

We make use of periodic boundary conditions. The lattice can then either represent an initially flat, topologically close and compact Universe, or a small region of a much larger classical patch.  

In any case, we consider an initial configuration of the Universe constituted of inhomogeneous scalar energy density $\langle \rho \rangle \gg \langle \rho_V\rangle$, compensated in the Einstein Equations by the scalar curvature of the metric. The initial energy density of vector and tensor modes of the metric is chosen to be zero (i.e. $\mathbf{\aleph} (\vec x) = 0$).

Below, we detail the methods used to solve the Hamiltonian constraint  and produce the chosen sets of ICs.

\subsection{Gradient-dominated initial conditions}

We consider different configurations for the initial inhomogeneities of the scalar field $\varphi$.
The Hamiltonian constraint (\ref{eq:HamBBSN}) is solved with an iterative method to obtain the corresponding initial distribution of the conformal factor $\psi$ on the lattice.
Like in previous works \cite{Clough:2016ymm, Aurrekoetxea_2020,  East:2015ggf}, the homogeneous value of the extrinsic curvature is chosen such that $K = -  \sqrt{ 24\pi\langle \rho \rangle}$.  Note that when calculating the average, the physical volume is dependent on the conformal factor $\psi$. The negative sign in $K$ reflects an initially expanding Universe.  The field velocity initially vanishes everywhere, and thus the momentum constraint (\ref{eq:MomBBSN}) is trivially satisfied.

The initial scalar field configuration is chosen as follows:
\be
\begin{split}
\varphi(t_0,\, & \vec x)  =  \bar\varphi_0  
                        + \Delta_e^\varphi \exp\left( \frac{ -(\vec x - \vec\mu)^2}{ \sigma^2  }\right)  \\
                        & + \frac{\Delta_{cos}^ \varphi}3  \left( \cos \frac{2\pi x}L + \cos \frac{2\pi y}L +\cos\frac{2\pi z}L \right)~.
\end{split}
\ee
Such a pattern can represent a Gaussian field fluctuation of amplitude $\Delta_{\rm exp}^\varphi$ at the centre of the lattice, on top of an inhomogeneous field value $\Delta_{\rm cos}^\varphi$. Here, $\vec\mu$ and $\sigma$ denote the mean and variance of the Gaussian mode. Unless otherwise specified, we choose $\vec\mu = (L/2, L/2, L/2),~ \sigma = L/6$, with $L$ being the physical size of the lattice.

\subsection{Kinetic-dominated initial conditions \label{sec::KDics}}

For this set of initial conditions, we fix a homogeneous scalar field by imposing
$\varphi (\vec x)  = \varphi_0$. The other ICs can be solved in three different ways:
\begin{enumerate}
\item
Analogously to what is done for gradient initial conditions, one can choose an initial inhomogeneous configuration for $\Pi_{\rm M}$ and solve equation (\ref{eq:HamBBSN}) to obtain $\psi$ by allowing a homogeneous $K$.  This method is prone to fail to converge due to the existence of non-unique solutions.
\item
By solving $\Pi_{\rm M}$ from an arbitrary configuration in the energy density, at given homogeneous field value $\varphi = \varphi_0$,
\begin{align}
    \Pi_{\rm M} &= \pm \sqrt{2(\rho - \rho_{V}) } \\
    \rho_V &= V(\varphi_0) = \rm {const.}
\end{align}
This method will often lead to initial conditions with only either positive or negative $\Pi_{\rm M}$ regions. 
\item
By setting an inhomogeneous conformal factor $\psi$ and then solving $\Pi_{\rm M}$ with
\be
    \Pi_{\rm M}^2 = - \frac{\psi^{-5}}\pi  \nabla^2\psi(x) +  \frac1{24} K^2(x) - \rho_V %
\ee
where $K$ here is given by
\begin{align}
    K &= \sqrt{ 24(\Psi + \rho_V)}\\
    \Psi &= \max \left( \frac{ \psi^{-5}}\pi \nabla^2\psi(x) \right)
\end{align}
In particular, we choose $\psi$ of the form  
\be
\begin{split}
& \psi = \exp\left(\zeta/2 \right) 
\\
\zeta = \exp& \left( - \vec x^2/\sigma_\zeta^2 \right) ~, \quad \sigma_\zeta  \approx \frac{L}{10}
\end{split}
\ee
which generates a spherical $\Pi_{\rm M}=0$ contour centred on the simulation grid,  allowing positive and negative values of $\Pi_{\rm M}$.
\end{enumerate}

\begin{figure*}[t!]
\begin{center}
\includegraphics[width=16cm]{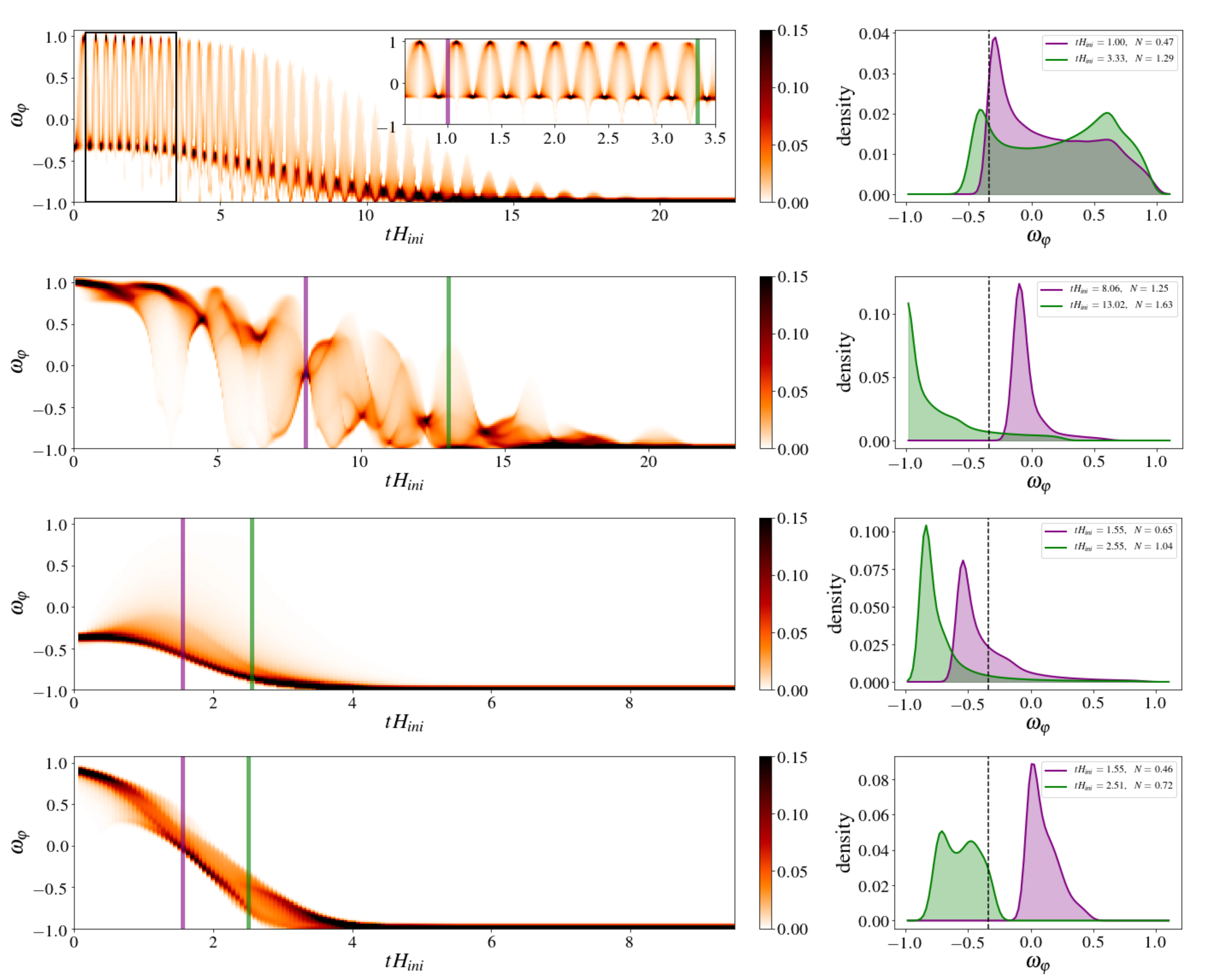}
\caption{ 
Illustrative examples of the dynamics of the equation of state for sub-Hubble and gradient and kinetic initial conditions (Top, centre-top), and super-Hubble with gradient and kinetic initial conditions (centre-bottom, bottom). On the left, the distributions on the physical grid are represented over Hubble times. Vertical purple and green lines within indicate selected times for which the distribution is represented by the right panels. Vertical dotted lines indicate the $\omega_\varphi = - 1/3$ threshold.
\label{fig:EOS_sample}}
\end{center}
\end{figure*}

\section{Results} \label{sec:results}

In this section, we present the main results of our simulations.  The paradigm that we consider is a universe dominated by a single scalar field, initially conformally flat, and in absence of ECMs.  Nonetheless, ECMs will develop during the time evolution and affect the dynamics of the system. Such modes are then   \textit{self-generated} by the inhomogeneities present during the evolution.
We also emphasize that in models with a plateau potential like Starobinsky or Higgs inflation, the effect of the potential on the highly inhomogeneous dynamics of preinflation is marginal and only provides the reference energy scale $H_{\rm {Inf}} $ at which inflation begins.  Because we chose a mean scalar field value on the slow-roll plateau, one has $H_{\rm {Inf}} \approx 10^{-6} M_{\rm pl}$.
\\
It is important to notice that $H_{\rm {Inf}}^{-1}$ \textit{is not} related to the size of the Hubble horizon prior to the inflation onset.  During preinflation, this depends on the total energy content in given volumes, and for our choice for the initial conditions, the initial Hubble scale is ${ H_{\rm ini} \approx \langle \rho \rangle^{1/2} }$, as illustrated in Figure \ref{fig:scale_diagram}.

By considering fluctuations of size similar to the lattice size $L$, one can distinguish between super-Hubble ($L > H_{\rm ini}^{-1}$) and sub-Hubble ($L \lesssim H_{\rm ini}^{-1}$) initial conditions.  Only the Planck scale limits the initial energy density, and thus the amplitude of field fluctuations.  In our simulations, we only consider energy densities up to two orders of magnitudes from it.

\subsection{Homogenisation phase}  

The scalar field gradients in principle decay like ${\rho_{\rm grad} \propto a^{-2}}$, while
its kinetic energy
scales like ${\rho_{\rm kin} \propto a^{-6}}$. However,
these two contributions are observed to alternatively dominate the total energy content, and shortly after the onset of the simulations, the effective scaling goes like
\be
\langle \rho_{\rm grad} +  \rho_{\rm kin} \rangle \propto a^{-4} ~ .
\ee
This mixing behaviour, to some extent, is generic for both gradient and kinetic dominated initial conditions, but this takes place at a slower rate in the case of a large and dominant background field velocity.  Illustrative examples of this evolution are provided in Figure~\ref{fig:EOS_sample}.

In the gravity sector, the breaking of the initial ``staticity" (allowing $S_i \neq 0$), triggers perturbations in the extrinsic curvature. This is manifested by a growing variance of its components, i.e. with ${\langle \Delta K \rangle ^2 = \langle \left(K - \langle K\rangle \right)^2 \rangle > 0}$, and similarly  $\langle \Delta \mathbf{\aleph} \rangle >0 $. After a short period of time, a new inhomogeneous equilibrium is reached and a  ``re-homogenisation" phase begins leading approximately to the following scaling relations during preinflation,  

\be
\begin{split}
\langle  K \rangle \propto \sqrt{\langle\rho \rangle} ~, \quad
\langle \Delta K \rangle \propto a^{-2} ~, \quad
\\
\langle \mathbf{\aleph}  \rangle
\propto a^{-2}  ~, \quad
\langle \Delta \mathbf{\aleph} \rangle \propto a^{-2} ~.
\end{split}
\ee

The dynamics vary depending on the particular example. Such complexity is due to the combination of source terms, and the Hubble friction present in Eq.~(\ref{eqn:dtAij2}). At early times, the contribution from the $R\ij^{TF}$-term is dominant, which initially corresponds to the spatial variation in the scalar curvature. 

\begin{figure}[t]
\begin{center}
\hspace*{-0.5cm}
\includegraphics[width=8.5cm]{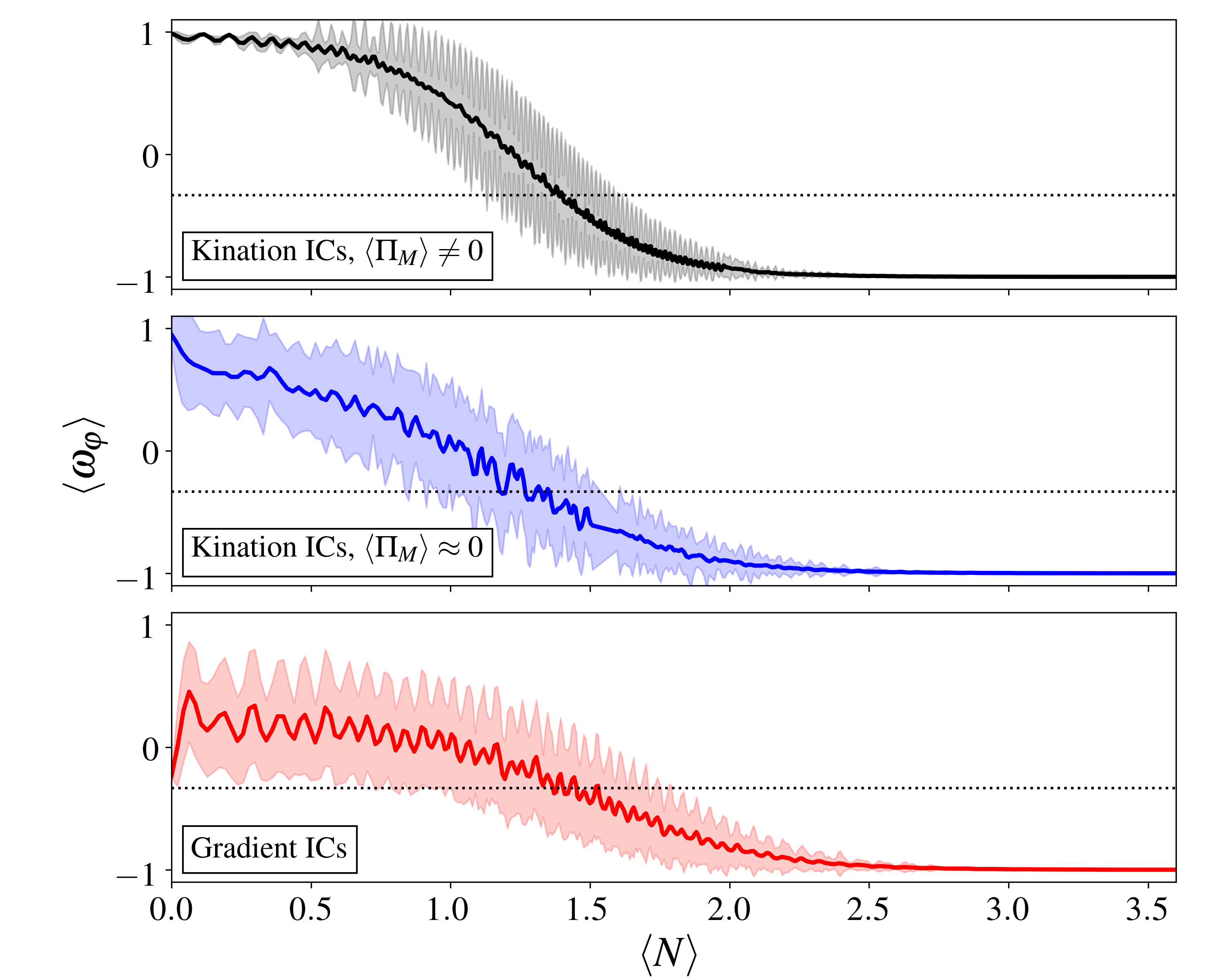}
\caption{ Mean equation of estate (solid line) and standard distribution (showed region) over mean efold of expansion. Black and blue lines represent cases with sub-Hubble Kinetic initial conditions, Red line corresponds to sub-Hubble initial conditions in the form of gradients. Dotted horizontal line marks the threshold $\omega_\varphi = -1/3$.
\label{fig:mean_EoS}}
\end{center}
\end{figure}

\begin{figure}[t]
\begin{center}
\hspace*{-0.5cm}
\includegraphics[width=8.5cm]{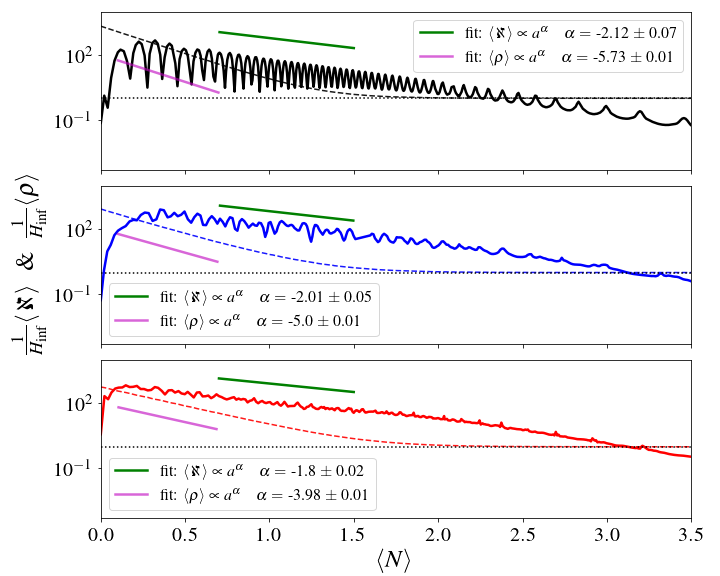}
\caption{ Same cases as in figure \ref{fig:mean_EoS}, but for means of the metric (solid line), and scalar field (dashed line) energy density. Green and pink lines correspond to linear fits of the scaling relations of the scalar field and metric energy densities, respectively.  In all plots, y-axis have been normalized over the inflationary scale $H_{\rm INF}$, marked in the plot with a horizontal dotted line.
\label{fig:mean_ECM}}
\end{center}
\end{figure}

\subsection{Oscillatory equation of state}

As we discussed in Section \ref{sec:cors_universe},
the onset of inflation, in the sense of accelerated expansion, is conditioned on the equation of state being $\langle\omega_\varphi \rangle < - 1/3$.
In the following, we show that the evolution of the equation of state, for all the different sets of initial conditions, generically leads to inflating regions.

For sub-Hubble modes, the top panel of Fig \ref{fig:EOS_sample} displays the time evolution of the distribution over the physical lattice of $\omega_\varphi$, for an illustrative case initially dominated by gradient energy density due to a single sinusoidal mode for the scalar field. It puts in evidence the oscillatory behaviour of  $\omega_\varphi$.  During this phase, the effective equation of state corresponds to a radiation dominated universe with  $\langle \omega_{\varphi} \rangle_{\rm eff}\approx 1/3$, which transits to potential domination and almost de-Sitter expansion, after roughly $\Delta N \approx 2$ efolds, where $\langle \omega_{\varphi} \rangle_{\rm eff} \approx -1$.  In this case, short-lasting contracting regions appear, whose integrated mass is above the Planck mass, but no black hole has been found by using an apparent-horizon finder code.

Conversely, for the super-Hubble gradient scenario shown in the third panel, the overall region tends to smoothly transition into the inflation, where most of the energy is found to be dissipated in a few Hubble times. However, part of the initial gradient energy is feeding a relativity small but strongly contracting region, undergoing kination, and forming a black hole (of mass $M_{\rm BH} \approx 10^5 M_{\rm pl}$).

Figure \ref{fig:EOS_sample} also shows examples of kinetic-dominated initial conditions.  In the second panel, we consider (large) sub-Hubble fluctuations in the initial field velocity on top of a non-vanishing value.  It shows a strongly varying equation of state. Long-lasting contracting regions appear after four Hubble times but still no black hole is found. The latter panel corresponds to a similar case but now initial kination at super-Hubble scales. The kinetic energy density is dissipated entirely with almost no mixing into gradients. This particular case smoothly transitions into inflation and without forming black holes.

\begin{figure}[b]
\begin{center}
\hspace*{-0.5cm}
\includegraphics[width=8.5cm]{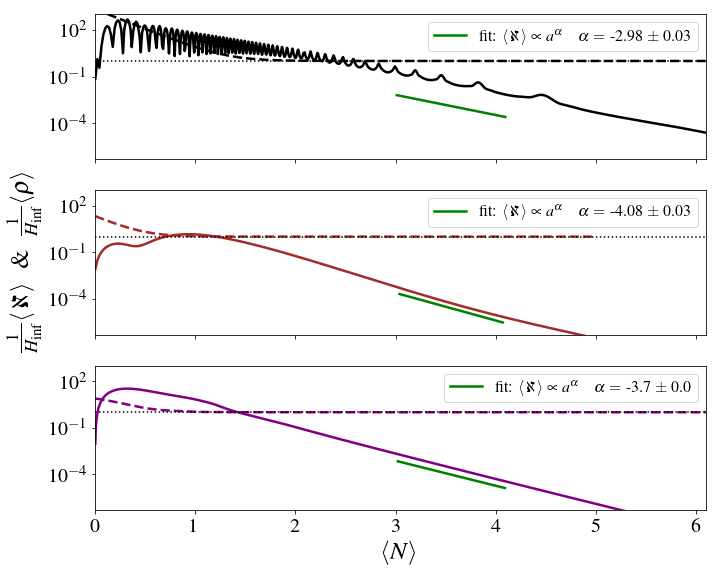}
\caption{ Similar to Fig. \ref{fig:mean_ECM}, but for late-time dynamics. The top  panel corresponds to sub-Hubble kination ICs, while the middle and bottom panels correspond to super-Hubble kination and gradient ICs, respectively.
\label{fig:mean_postinf_ECM}}
\end{center}
\end{figure}

Additional cases are displayed in Figure \ref{fig:mean_EoS}, which shows the mean evolution of the equation of state with sub-Hubble inhomogeneities. The first two examples respectively correspond to kination with and without a dominant background, and the third one corresponds to a configuration with gradients. Remarkably, the dynamics of mean averages of inhomogeneous kination are in very good agreement with the semi-analytical solution of the \textit{homogeneous} case (e.g. see Sec. II-C in Ref. \cite{Chowdhury:2019otk}).
Figure~\ref{fig:mean_EoS} can be complemented with Figure \ref{fig:mean_ECM}, where the mean metric and scalar field energy densities are displayed for the same examples (see colour-code).

\begin{figure*}[t]
\begin{center}
\hspace*{-2mm}
\includegraphics[width=8.5cm]{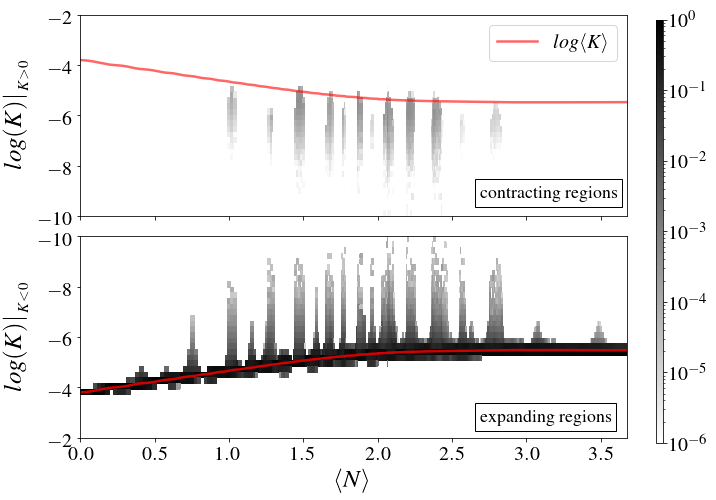} \hspace*{4mm}
\includegraphics[width=8.5cm]{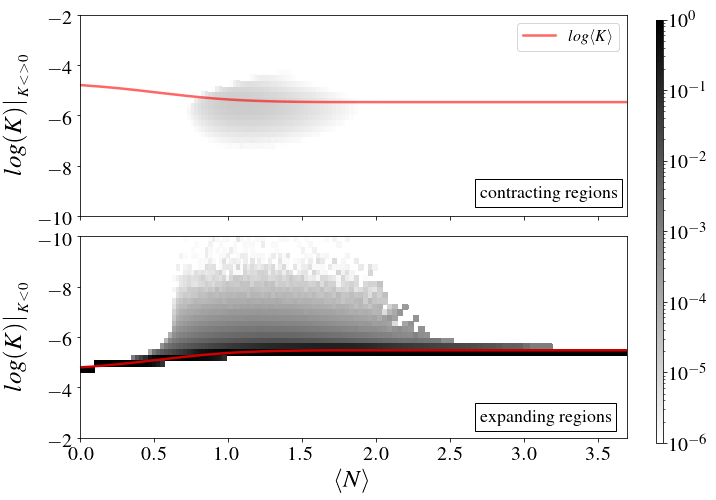}
\caption{ Distribution (logarithmic scales) of $|K|$ on the physical grid over mean number of efolds,  divided into contracting (top) and expanding regions (bottom). The red solid line indicates the mean of $K$. The plots corresponds to the simulations with sub-Hubble gradient initial conditions (left) and  with super-Hubble gradients (right) shown in Fig. \ref{fig:EOS_sample}. For the super-Hubble case, the formation of a black hole with mass $ M_{BH} \approx 2.6 \cdot 10^4 M_p$ is confirmed. After $N\approx 2$, the black hole falls down from the resolution grid of our simulation and we lose track of it. 
\label{fig:subHub_BH}
}
\end{center}
\end{figure*}

\subsection{Generation of extrinsic curvature modes}

The observed dynamics in the equation of state does not go unnoticed in the gravity sector.  
We find that ECMs are generated in the case of (large) sub-Hubble scalar field inhomogeneities.  This is not the case for super-Hubble fluctuations, for which there is no oscillation between the energy in the metric and the other energy components.  Instead the metric energy smoothly grows in contracting regions that end up forming black holes. The rise of $\langle \mathbf{\aleph}\rangle$ can therefore be seen as a byproduct of preinflation black hole formation.  But since this happens on super-Hubble scales, they do not have a significant impact on the expansion dynamics elsewhere.  \\

For oscillatory sub-Hubble ECMs the situation is different. These modes are strongly sourced at early times as a product of non-vanishing anisotropic stress tensor and Ricci tensor. The $\langle \mathbf{\aleph} \rangle$ reaches a maximal mean value in less than $\Delta N\approx 0.5$ efolds, and then it decays approximately as $\langle \mathbf{\aleph} \rangle \propto a^{-2}$ during the rest of preinflation (see Fig. \ref{fig:mean_ECM}).  Because the decay rate of ECMs is lower than for other energy components,
they can potentially provide, on average, a dominant contribution to the preinflation dynamics. Nonetheless, this scenario is restricted to very large sub-Hubble perturbations in the matter sector. One typically needs dominant perturbation sizes of $ \lambda/ H^{-1} \gtrsim 0.1 $  for the effect to be significant. Yet, this condition is weaker at higher energy scales ($\langle \delta\rho \rangle \sim 0.01 ~ M_{\rm pl}$), because of the longer duration of the homogenisation phase and given that $\langle \mathbf{\aleph} \rangle / \langle \rho \rangle \propto a^2$.

In Fig. \ref{fig:mean_postinf_ECM} the late-time scaling is shown, once accelerated expansion has begun. During the transition towards inflation, their decaying rate gradually strengthen, reaching up to $\langle \mathbf{\aleph} \rangle \propto a^{-4}$. For sub-Hubble ECMs, the oscillation frequency also gradually decreases, until they eventually ``freeze out" at horizon exit.

\subsection{Contracting regions and black holes}

Even if the condition $\langle K \rangle < 0$ is always satisfied,
overdensities may generate local contracting regions, themselves embedded in expanding ones. 
Figure \ref{fig:subHub_BH} shows examples of sub-Hubble and super-Hubble initial conditions, where time-histograms of expanding and contracting regions are plotted in the top and bottom panels, respectively. Contracting regions may develop because of overdensities in either the scalar field or metric energy densities.  

When they occur because of the matter sector, they are formed inside kinetic dominated regions driven by strong scalar field Laplacians that overcome the Hubble friction in Eq.~({\ref{eq:dtPi}). Once contraction is triggered, the Hubble friction turns into a kind of Hubble boost for the kinetic energy, enhancing the formation of black holes.

When contracting regions develop because of overdensities in the metric sector, they originate from the ECMs sourced by the sub-Hubble scalar field dynamics. 
The formation of such contracting regions allows us to speculate on a distinct mechanism for (low-mass) black hole production,  consisting of the gravitational collapse of ECMs in the sub-Hubble regime. 
However, we have not been able to confirm this by finding the apparent horizon with our code. This could happen when the radius of the forming black hole is smaller than the resolution of the grid. Another possibility is that black hole formation is aborted if the equation of state rapidly approaches the inflationary attractor {$\omega_\varphi \approx -1$}. Determining whether or not these contracting regions develop into black holes is left for a future study.

\begin{figure}[b]
\begin{center}
\includegraphics[width=8.5cm]{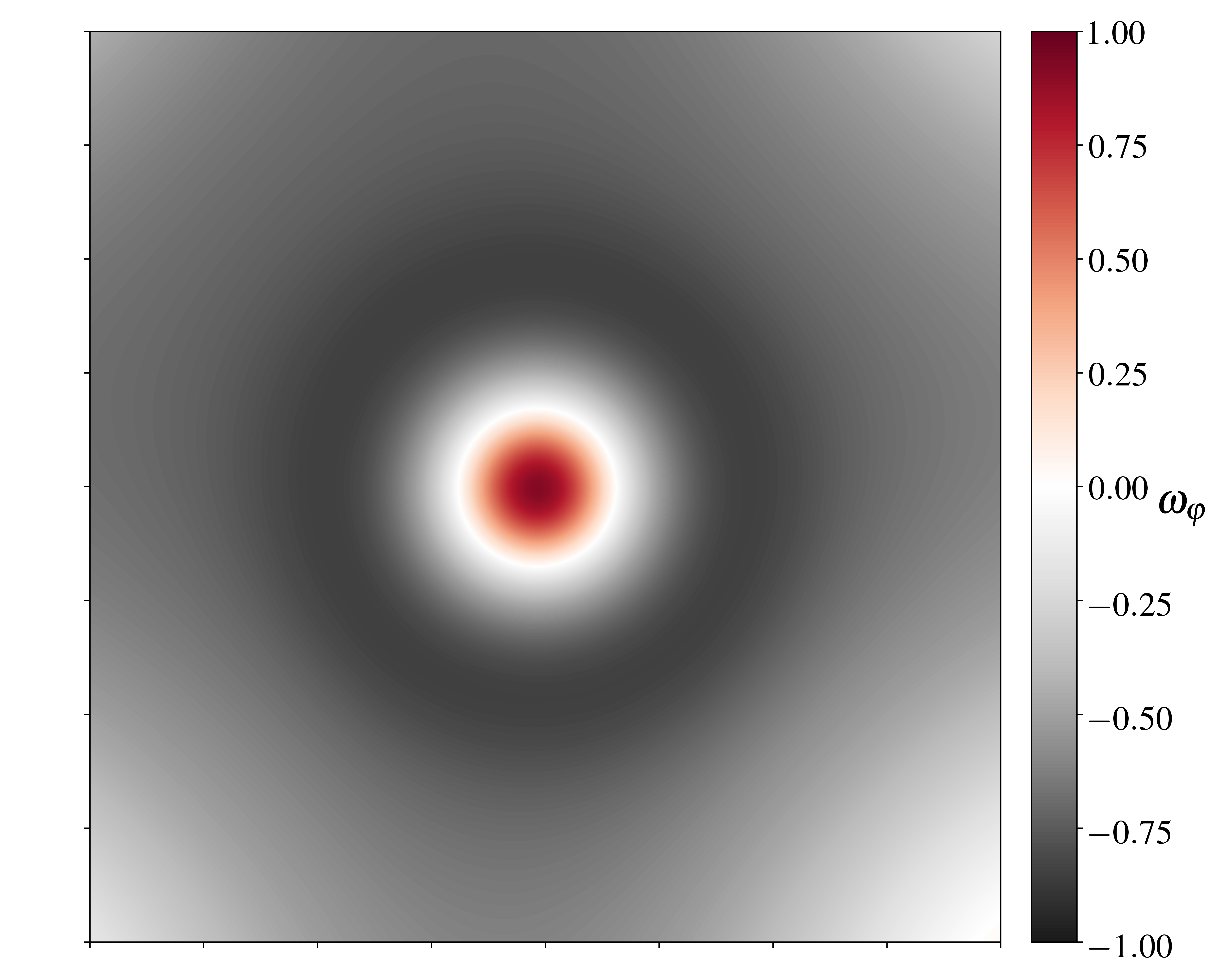}
\caption{
Equation of state in the black hole equatorial slice, corresponding to the example shown in Figure \ref{fig:subHub_BH} (right)  at $N \simeq 1$.  The central red disk corresponds to the kination region and coincides with the region where the black hole forms.  The surrounding darker shell is a region where the equation of state is close to ${\omega_\varphi=-1}$, and thus to the inflationary attractor. This suggests that the PIBH formation acts as a catalyst for inflation. }
\label{fig:BH_slc}
\end{center}
\end{figure}

For scalar field overdensities on super-Hubble scales, the above described kination-trigger mechanism occurs, and we confirm the results of previous works reporting on the formation of PIBHs \cite{Clough:2016ymm}. 
An apparent horizon is found and so the black hole mass can be inferred. In our simulations, the mass range of such black holes varies from $10$ to $10^5 ~M_p$.  Like in Ref. \cite{Clough:2016ymm}, the PIBH mass saturates at a given perturbation threshold, and decreases for very high amplitudes of the perturbations.  
It is also interesting to simulate the black hole formation with slightly more rigid gauge drivers, in order to have an alternative picture of the collapsing region. In Fig. \ref{fig:HExt_Sample} we provide such an example where, starting from super-Hubble gradient-dominated initial conditions, we find that a bifurcation in the energy density between the expanding and collapsing region occurs after $N \sim 0.6$ efold. Interestingly, we notice that homogenisation is enhanced in the neighbourhood of the collapsing region, due to the gravitational pull of the yet-to-form black hole.  This suggests that super-Hubble scalar field fluctuations, while forming PIBHs, may also facilitate the onset of inflation due to a higher level of homogenisation around the black hole, which also ``traps'' most of the generated metric energy.

\section{Discussion on preinflation} \label{sec:discussion}

The results of our lattice simulations have a series of implications  (discussed below) about the naturalness of (chaotic) inflation, the emergence of classical conditions from a suspected quantum-gravity regime, the dynamics and duration of the preinflation era, the formation of preinflationary black holes.

\subsection{On the universal dynamics}

Our analysis extends previous works by considering large inhomogeneities not only in field gradients but also in the field velocity.  In both cases, the simulations for sub-Hubble fluctuations have produced similar dynamics for the preinflation era, with oscillations between kinetic and gradient terms in the density.  This result reinforces the robustness of inflation to its initial conditions also in the case of non-linear sub-Hubble inhomogeneities.  Even if our initial conditions still assume a homogeneous expansion rate and conformal flatness, this configuration quickly evolves towards more general inhomogeneous configurations for all BSSN variables, including tensor and vector modes in the metric.  Therefore, our simulations also suggest that such behaviour is universal and insensitive to the exact configuration of the initial conditions.  
\\

Nevertheless, we have also identified a regime named \textit{kinaton}, in which the inhomogeneity in the field velocity are on top of a background velocity.  In such a case, the density is dominated by the kinetic term and the effective equation of state remains close to $\omega_\varphi=1$.  One can therefore wonder if drastic inhomogeneous configurations for the extrinsic curvature, together with gradients and kinetics in the scalar field, could also lead to radically different regimes.  This problem is left for future work, because numerically solving the Hamiltonian and momentum constraints in such a case is still a challenge \cite{Garfinkle_2020}.

\subsection{ On the characteristic scale of the inflaton} 
The Hubble scale is one of the fundamental notions in order to understand gravitational interactions at large scale, because modes freeze at super-Hubble distances. However, in inhomogeneous scalar field cosmologies the Hubble scale depends, in part, on the contributions of gradient and kinetic energies to the total energy density. Therefore, assuming a universe dominated by gradients, the largest sub-Hubble modes (i.e. those within the realm of classical Gravitation) are bounded.  Indeed, by roughly identifying the gravitational energy with 
\be
\rho_{\rm grad} \approx  \frac 12 \left( \frac{ \delta\varphi}{ \lambda}\right)^2~,
\ee
where $\delta\varphi \equiv \varphi_{\rm max} -\varphi_{\rm min}$ is the allowed scalar field variation and $\lambda$ the mode wavelength of size 
$\lambda = H^{-1}_{\rm ini}$, and by means of the usual Friedman equation, one finds that 
\be
 \delta\varphi  \lesssim  \sqrt\frac 3 {4\pi}  \approx  \frac12 ~ m_{\rm p}~.
\ee 

For plateau-like models where the field difference between the slow-roll region and the bottom of the potential is super-Planckian,
the potential gradient is in general small.  The average value of the field follows the Klein-Gordon equation with a strong friction term $H_{\rm ini}$, and almost negligible driver $V'(\varphi)$ everywhere. Thus, during the preinflation the trajectory of the mean scalar field down the potential is strongly suppressed, ensuring a large number of efolds after inflation begins. This is not the case for small-field potentials along the sub-Planckian slow region \cite{Clough_2015, Aurrekoetxea_2020}. This is the same reason why large-field models are in general more robust to the initial conditions than the small-field ones. 
\\

A similar reasoning can be used when the field has inhomogeneous kinetic initial conditions, when kinetic and gradient terms mix and mimic a radiation-domination era.  In the case of a dominant background (super-Hubble) field velocity, the system is analogous to the homogeneous kination phase discussed in Ref. \cite{Chowdhury:2019otk}, and in agreement with our simulations it shows that Starobinsky inflation produces a sufficient number of efolds during inflation when starting at super-Planckian field values, i.e. $\langle \varphi\rangle_{\rm ini} \gtrsim m_{\rm p}$.

~ \\
\subsection{On the preinflationary black hole formation}

The transition from gradient- to kinetic-dominated energies can give rise to contracting regions, typically where the scalar field Laplacian is the largest.  When this happens, the equation of state is $\omega_\varphi \approx 1$, one has locally
\be
-K < 0, \hspace{1cm} \partial_t (-K) < 0
\ee 
and there is simply no possibility for this region to expand again.  Therefore, this situation generically leads to the formation of PIBHs and the use of the puncture gauge guarantees that the lattice only probes the region surrounding those black holes.  For sub-Hubble fluctuations, there are many oscillations, and we observe the formation of a contracting region at similar locations.
If these regions form smaller PIBHs, those oscillations can thus either feed previously formed black holes, which would then grow in mass until inflation takes place, or  produce new PIBHs.  Their mass is of the order of the Planck mass for sub-Hubble fluctuations, so they would evaporate relatively quickly and produce a particle bath that is not taken into account in simulations, as well as eventual Planck relics.  In any case, the subsequent phase of inflation dilutes them such that the density of these relics in our Hubble volume would be extremely small or vanish.

\subsection{On the duration of the preinflation era}

\begin{figure*}[t]
\begin{center}
\hspace*{-0.5cm}
\includegraphics[width=18cm]{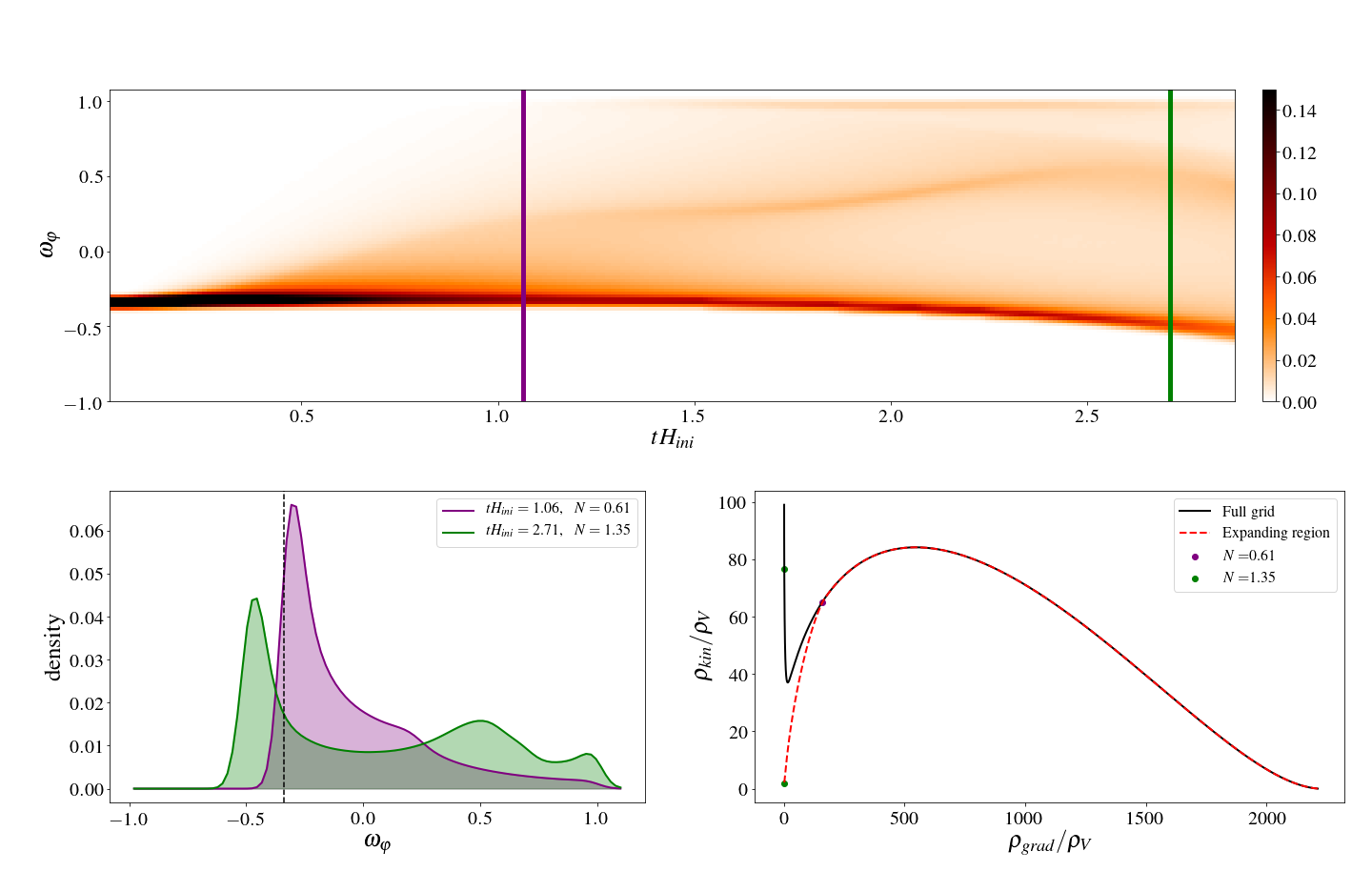}
\caption{ The top and bottom-left panels are similar to figure \ref{fig:EOS_sample}, but for a critical simulation with super-Hubble gradient initial conditions evolved shortly after a black hole is formed in the $\omega_\varphi \approx 1$ region. The bottom-right plots indicates trajectories in terms of the means of  kinetic and gradient energies, both normalized over the potential. The solid-black line corresponds to the whole grid averaging, while the dotted-red line only averages on expanding regions. 
\label{fig:HExt_Sample}}
\end{center}
\end{figure*}

A common picture of the very early Universe is that of a classical phase, described by General Relativity, emerging from an unknown regime of quantum gravity at the Planck scale.  During this phase, there is no reason supporting homogeneity on scales larger than the initial Hubble radius.  For a while, this was considered as a potential difficulty for inflation.  But simulations in full numerical relativity, like the ones presented in this work, show that non-linear Hubble-sized and sub-Hubble inhomogeneities do not prevent the onset of inflation after the density is damped by expansion at the level of the potential energy of the scalar field.  Several regimes have been identified for the preinflation era, characterized by different effective equation of state. We have emphasized the importance of  the $\mathbf{\aleph}$-term in the dynamics of the expansion.

This allows to set a limit on the maximal duration of the preinflation phase, expressed in terms of an averaged number of e-folds of expansion $\Delta N_{\rm preinf}$.  If the energy stored in field gradients, damped like $\rho_{\rm grad} \propto  \exp(-2N)$, were dominant throughout preinflation, one would get  $\Delta N_{\rm preinf} \approx - \ln \left( \rho_{\rm V}\right) /2$.  However, because of the wobbling between kinetic and gradient contributions, the density is effectively damped like ${\rho \propto \exp(-4N)}$ on average, and thus $\Delta N_{\rm preinf} \approx \log \left( \rho_{\rm V}\right) / 4 \simeq 7 $, for Higgs or Starobinsky inflation.

If one now allows the energy to be stored in the form of ECMs, preinflation can last up to $\Delta N_{\rm preinf} \approx 14$. These numbers increase if inflation took place at lower energy, and inversely.  It can also be reduced if the energy scale of quantum gravity is lower than the Planck scale.

\subsection{On the possible observable signatures and the minimum size of the Universe}

In the case where the time of Hubble exit of cosmological scales occurs only a few e-folds (denoted $N_{\rr i*}$ ) after the onset of inflation, preinflation should leave observable signatures in the CMB temperature anisotropies and polarization.  Indeed, in such a case the gradient and kinetic energies are not damped well below the scalar field potential energy and still slightly impact the expansion rate during inflation.  The $\aleph$-term can also have a similar effect.  This modifies the evolution of the Hubble-flow parameters
\be
\epsilon_1 \equiv - \frac{\rr d \ln H}{\rr d N}, \hspace{1cm} \epsilon_2 \equiv \frac{\rr d \ln \epsilon_1}{\rr d N}~,
\ee
and in turn modifies the predicted scalar spectral index $n_{\rm s}$ and tensor-to-scalar ratio $r$,
\be
n_{\rm s} = 1 - 2 \epsilon_1 - \epsilon_2, \hspace{1cm} r = 16 \epsilon_1~,
\ee
at first order in slow-roll parameters. This certainly requires a significant amount of tuning for these effects to be observable, allowing for just enough inflation, without totally spoiling the (almost) scale invariance of the primordial scalar power spectrum.  

Indeed, the dynamics of the expansion rate is not only driven by the potential and the kinetic energy of the field, but is also somehow impacted by gradient and metric terms.  One has to consider an effective energy density $\rho_{\rm eff} = \rho_V + \delta \rho + \aleph / 16 \pi$, and take $\delta \rho + \aleph / 16 \pi \approx \rho_V \exp(-\eta N_{\rm i*})$ after the onset of inflation, with the effective value of $\eta$ depending on which preinflation regime the ICs belong to.  If the dynamics of $\epsilon_1$ is dominated by these terms coming from the preinflaiton era, one gets $ \epsilon_1 \approx \eta \exp(- \eta N_{\rr i*} )$ and $\epsilon_2 = - \eta^2 $.  An important fine-tuning of the initial conditions of the scalar field is therefore required to get a value of $N_{\rr i*}$ that can explain the observed scalar spectral index.  Furthermore, this would unavoidably generate a tensor-to-scalar ratio $r \simeq 16 \epsilon_1$ larger than unity, which is excluded. 

If one assumes that the dynamics of $\epsilon_1$ is governed by the potential, which gives $\epsilon_1 \simeq 2 \times 10^{-4}$ for Higgs/Starobinsky inflation, gradient or metric terms may still govern the dynamics of $\epsilon_2$.  One then gets $\epsilon_2 \approx - \eta^2 \exp(-\eta N_{\rr i*}) / \epsilon_1 $. 
As a consequence, the preinflation era could have an observable effect on the primordial power spectrum in a \textit{just-enough} inflation scenario and could even give the correct value of the scalar spectral index, at the price of a significant fine-tuning of the initial scalar field.  If the effects on the second Hubble-flow parameter (which is determined at the $20\%$ level) are not detectable, one gets from the current limits on the tensor-to-scalar ratio a lower bound $N_{\rr i*} \gtrsim 3 $. An interesting corollary is that, in the above-mentioned scenario, the non-detection of the imprints of preinflation implies that the Universe is at least $\exp[3 (N_{\rm preinf}+ N_{\rm i*})] \gtrsim 10^{13}$ times larger than our current Hubble volume.

\subsection{Is Linde's picture right?} 

In all of the considered cases and regimes inflation is a generic outcome, at least in most parts of the lattice. This conclusion applies to all expanding (on average) initial conditions with scalar field fluctuations of any size, as long as there exists a Hubble patch whose mean-field value is in the slow-roll region of the potential.
On the contrary, inflation cannot be triggered from sub-Hubble fluctuations around the bottom of the potential, even if apparently the energy density initially stored in field gradients or velocity is much larger than the potential barrier to reach the slow-roll region.  The reason is that field gradients act as a damping term in the Klein-Gordon equation for the scalar field while large gradients in the field potential ``drag" the scalar field down towards the bottom, as seen in Ref. \cite{Aurrekoetxea_2020}.
One should not interpret this as an issue, since a large background field value may have emerged in an initially flat and compact Universe, with a 3-torus topology, as suggested by Linde \cite{Linde:2004nz,Linde:2017pwt,Linde:1983gd,Goncharov:1983mw,Linde:1983mx,Kofman:1985aw}, in support of chaotic inflation.  The second possibility is a much larger inhomogeneous Universe, in which it is sufficient to have a single super-Hubble patch with a large mean-field value superimposed by arbitrary sub-Hubble fluctuations, to naturally lead to inflation.  
\\

One could also wonder how the topology, the initial curvature, and shape of the Universe could influence preinflation and even prevent inflation.  Despite that we have been using periodic boundary conditions on a cubic lattice, reflecting the topology of a 3-torus, in the presence of large inhomogeneities the Universe is formed by locally open and closed regions as explained in Ref. \cite{East:2015ggf}. Therefore, even if a homogeneous closed Universe would collapse when the energy content is dominated by curvature \cite{Linde:2004nz, Linde:1987aa}, in the largely inhomogeneous case local open regions may still be expanding and eventually lead to inflation. 
However, scenarios with globally large positive curvature have not been tested yet. This is left for future work.

\section{Conclusions}  \label{sec:ccl}

Fully relativistic lattice simulations in 3+1 dimensions provide the ideal method to study the possible non-linear inhomogeneous dynamics of the preinflation era.   In this work, we have extended previous analyses and considered new realizations of the initial conditions satisfying the Hamiltonian and momentum constraints, with an inhomogeneous scalar field velocity.  Even if in general the preinflation era does not leave distinguishable signatures in observations, it is related to fundamental questions such as how natural or fine-tuned are the initial conditions of inflation and how generic is the transition from a semi-classical regime emerging from some quantum gravity period, towards inflation.  In particular, the realisation of inflation starting from non-linear and sub-Hubble field fluctuations has been a long-standing and debated problem.

Besides confirming recent results on the robustness of inflation to sub-Hubble inhomogeneities, our simulations revealed richer preinflation dynamics than expected, as well as some universal behaviours.  A new regime in which the expansion is driven by the traceless part of the extrinsic curvature tensor has been identified, which impacts the equation of state and the duration of the preinflation era.  We have also found that initial conditions with highly inhomogeneous scalar field velocity give rise to a regime in which the density remains dominated by the kinetic term, denoted \textit{kination}.  Otherwise, sub-Hubble and Hubble-sized inhomogeneities give rise to oscillations between gradient- and kinetic-dominated periods.  For super-Hubble sized fluctuations, our analysis also confirms the emergence of contracting regions, leading to the formation of preinflation primordial black holes, those being subsequently diluted by inflation.  Our findings are summarized in Fig.~\ref{fig:scale_diagram} where we show the possible outcomes starting from different regions of the parameter space.

The chosen set of initial conditions, even if extended to highly inhomogeneous field velocities, still corresponds to very specific cases.  Indeed, the initial extrinsic curvature tensor is taken with a vanishing traceless part $A_{ij}$ and a homogeneous negative trace $K$.  Nevertheless, we point out that for Hubble-sized and sub-Hubble inhomogeneities, the extrinsic curvature tensor rapidly becomes highly inhomogeneous, thereby exploring various other configurations, until it eventually drives the overall expansion dynamics.  Our initial data also correspond to a Ricci curvature that remains small on average (but it can be very large locally).   But the issue of solving the momentum and Hamiltonian constraints on the initial hypersurface for more general cases remains a major computational challenge that still limits the range of applicability of the simulations of the preinflation era in full 3+1 numerical relativity.} We also did not consider the eventual impact of additional scalar of matter fields on the preinflation dynamics, or other topological choices than periodic boundary conditions.  

Finally, we have focused on the Higgs - Starobinsky inflation model and the phenomenology of the preinflation era is marginally impacted by the field potential, which only gives a sub-dominant contribution to the energy density during most of the preinflation phase. The potential only starts to dominate just before the onset of inflation and, therefore, our conclusions should remain valid for any plateau-type scalar field potential at super-Planckian characteristic values, these being currently favoured by CMB observations.  

In summary, our work contributes to paving the way to a better and more precise understanding of the rich phenomenology of the preinflation era.  It enlarges the diversity of initial conditions that have been considered so far.  But future work and the development of more advanced numerical methods will be needed to go beyond the assumption of initial conformal flatness, and to include inhomogeneous initial configurations for the extrinsic curvature in both $K$ and $\tA\ij$.  

Looking forward, we may also consider the effect of additional scalar or matter fields, analyze the dynamics for small-field potentials, and study more deeply the formation process of PIBHs.  In a semi-classical description of the preinflation era, it may also be possible to consider the non-linear effects of quantum scalar field fluctuations by adding a stochastic term in the Klein-Gordon equations. It may also be interesting to study the case where preinflation leaves observable signatures in CMB observations, even if this requires a significant fine-tuning.

Finally, our work emphasizes that in absence of these signatures and assuming that classical preinflation emerged at the Planck scale, the Universe is at least ten thousand billion times larger than our current Hubble volume, usually referred to as our observable Universe.  
\\

\section{Acknowledgments}

The authors warmly thank Christophe Ringeval, Andrei Linde,  Vincent Vennin, Julien Lesgourgues, Josu C. Aurrekoetxea, Katy Clough and Eugene Lim for useful discussions and support, as well as the \texttt{GRChombo} team (http://grchombo.github.io/collaborators.html) for their work on the code. 
The work of CJ is supported by the {FRIA} Grant No.1.E.070.19F of the Belgian Fund for Research, F.R.S.-FNRS. 
SC is supported by the \textit{Charge de Recherche} grant of the Belgian Fund for Research, F.R.S.-FNRS. 
Computational resources have been provided by the Consortium des Équipements de Calcul Intensif (CÉCI), funded by the Fonds de la Recherche Scientifique de Belgique (F.R.S.-FNRS) under Grant No. 2.5020.11 and by the Walloon Region.
Parallel code developments were done on the CURL cosmo clusters at UCLouvain, funded by the “Fonds de la Recherche Scientiﬁque - FNRS” under Grant No. T.0198.19. 
Analysis and visualizations employed the Visit \cite{10.5555/2422936} and yt-project \cite{Turk_2010} software packages. 
Finally, the authors acknowledge the popular BBC's \textit{Doctor Who} episode, ``Blink", which prompted the incredible sing-songy title (for the arXiv reader) of this work.  

\begin{figure*}[t!]
\begin{center}
\hspace*{-0.5cm}
\includegraphics[width=6.cm]{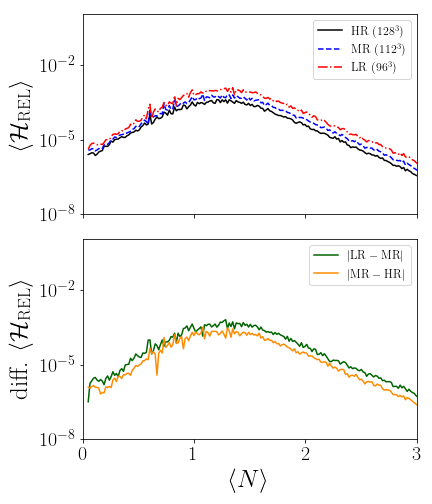}
\includegraphics[width=6.cm]{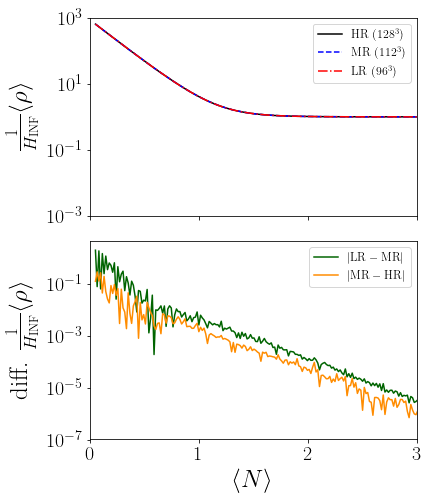}
\includegraphics[width=6.cm]{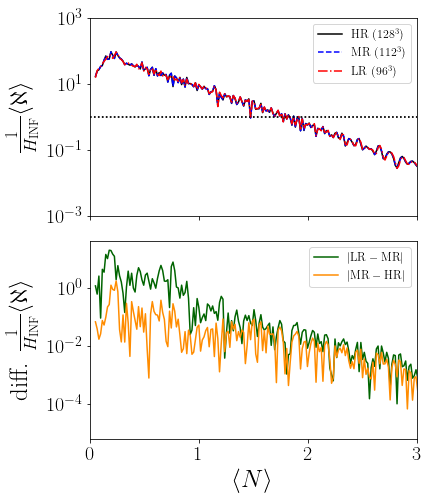}
\caption{ 
Convergence tests on the Hamiltonian constraint (left), mean scalar field energy density (center) and mean metric energy density (right).   Top panels:  evolution of the mean values for low (LR, red), medium (MR, blue) and high (HR, black) resolutions.  Bottom panels:  LR-MR (green) and MR-HR (orange) comparisons.
\label{fig:A2_convergence}
}
\end{center}
\end{figure*}

\appendix
\section{Code validation and convergence tests}

Besides ensuring that violations on the constrain equations (\ref{eq:HamBBSN}, \ref{eq:MomBBSN}) are always small, a common validation method is checking for numerical convergence when using different grid resolutions $\Delta x$. Because we use a fourth-order finite difference stencil when computing the gradients, we expect numerical errors to be degraded like $(\Delta x)^{4}$. In our convergence test, we run simulations with low (LR: $96^3$), medium (MR: $112^3$) and high (HR: $128^3$) resolutions grids. 
\\

In Fig. \ref{fig:A2_convergence}, taking an example from sub-Hubble kinetic ICs of Figs. \ref{fig:mean_EoS} and \ref{fig:mean_ECM}, we show the convergence test for the Hamiltonian constrain,  $\langle\rho\rangle$ and $\langle\aleph\rangle$.  
We have defined  
\be
  \mathcal{H}_{\rm REL} = \frac {\mathcal{H} }{\left[\mathcal{H}\right] }  
\ee
where 
\be
\left[\mathcal{H}\right]  \equiv \Biggl[  \lp {R} \rp ^2 +  \lp \tilde A\IJ \tilde A\ij \rp ^2 
+ \lp  \frac23 K^2 \rp ^2  + \lp 16\pi \rho \rp ^2 \Biggr] ^{1/2}
\ee

denotes the relative violations of the Hamiltonian constraint during the evolution. 
Convergence test plots are consistent with fourth-order convergence.

\newpage
\bibliographystyle{apsrev4-1}
\bibliography{biblio.bib}

\begin{thebibliography}{55}%
\makeatletter
\providecommand \@ifxundefined [1]{%
 \@ifx{#1\undefined}
}%
\providecommand \@ifnum [1]{%
 \ifnum #1\expandafter \@firstoftwo
 \else \expandafter \@secondoftwo
 \fi
}%
\providecommand \@ifx [1]{%
 \ifx #1\expandafter \@firstoftwo
 \else \expandafter \@secondoftwo
 \fi
}%
\providecommand \natexlab [1]{#1}%
\providecommand \enquote  [1]{``#1''}%
\providecommand \bibnamefont  [1]{#1}%
\providecommand \bibfnamefont [1]{#1}%
\providecommand \citenamefont [1]{#1}%
\providecommand \href@noop [0]{\@secondoftwo}%
\providecommand \href [0]{\begingroup \@sanitize@url \@href}%
\providecommand \@href[1]{\@@startlink{#1}\@@href}%
\providecommand \@@href[1]{\endgroup#1\@@endlink}%
\providecommand \@sanitize@url [0]{\catcode `\\12\catcode `\$12\catcode
  `\&12\catcode `\#12\catcode `\^12\catcode `\_12\catcode `\%12\relax}%
\providecommand \@@startlink[1]{}%
\providecommand \@@endlink[0]{}%
\providecommand \url  [0]{\begingroup\@sanitize@url \@url }%
\providecommand \@url [1]{\endgroup\@href {#1}{\urlprefix }}%
\providecommand \urlprefix  [0]{URL }%
\providecommand \Eprint [0]{\href }%
\providecommand \doibase [0]{http://dx.doi.org/}%
\providecommand \selectlanguage [0]{\@gobble}%
\providecommand \bibinfo  [0]{\@secondoftwo}%
\providecommand \bibfield  [0]{\@secondoftwo}%
\providecommand \translation [1]{[#1]}%
\providecommand \BibitemOpen [0]{}%
\providecommand \bibitemStop [0]{}%
\providecommand \bibitemNoStop [0]{.\EOS\space}%
\providecommand \EOS [0]{\spacefactor3000\relax}%
\providecommand \BibitemShut  [1]{\csname bibitem#1\endcsname}%
\let\auto@bib@innerbib\@empty
\bibitem [{\citenamefont {Akrami}\ \emph {et~al.}(2020)\citenamefont {Akrami},
  \citenamefont {Arroja}, \citenamefont {Ashdown}, \citenamefont {Aumont},
  \citenamefont {Baccigalupi}, \citenamefont {Ballardini}, \citenamefont
  {Banday}, \citenamefont {Barreiro}, \citenamefont {Bartolo},\ and\
  \citenamefont {et~al.}}]{Akrami:2018odb}%
  \BibitemOpen
  \bibfield  {author} {\bibinfo {author} {\bibfnamefont {Y.}~\bibnamefont
  {Akrami}}, \bibinfo {author} {\bibfnamefont {F.}~\bibnamefont {Arroja}},
  \bibinfo {author} {\bibfnamefont {M.}~\bibnamefont {Ashdown}}, \bibinfo
  {author} {\bibfnamefont {J.}~\bibnamefont {Aumont}}, \bibinfo {author}
  {\bibfnamefont {C.}~\bibnamefont {Baccigalupi}}, \bibinfo {author}
  {\bibfnamefont {M.}~\bibnamefont {Ballardini}}, \bibinfo {author}
  {\bibfnamefont {A.~J.}\ \bibnamefont {Banday}}, \bibinfo {author}
  {\bibfnamefont {R.~B.}\ \bibnamefont {Barreiro}}, \bibinfo {author}
  {\bibfnamefont {N.}~\bibnamefont {Bartolo}}, \ and\ \bibinfo {author}
  {\bibnamefont {et~al.}},\ }\href {\doibase 10.1051/0004-6361/201833887}
  {\bibfield  {journal} {\bibinfo  {journal} {Astronomy \& Astrophysics}\
  }\textbf {\bibinfo {volume} {641}},\ \bibinfo {pages} {A10} (\bibinfo {year}
  {2020})}\BibitemShut {NoStop}%
\bibitem [{\citenamefont {Ade}\ \emph {et~al.}(2016)\citenamefont {Ade},
  \citenamefont {Aghanim}, \citenamefont {Arnaud}, \citenamefont {Arroja},
  \citenamefont {Ashdown}, \citenamefont {Aumont}, \citenamefont {Baccigalupi},
  \citenamefont {Ballardini}, \citenamefont {Banday},\ and\ \citenamefont
  {et~al.}}]{Ade:2015lrj}%
  \BibitemOpen
  \bibfield  {author} {\bibinfo {author} {\bibfnamefont {P.~A.~R.}\
  \bibnamefont {Ade}}, \bibinfo {author} {\bibfnamefont {N.}~\bibnamefont
  {Aghanim}}, \bibinfo {author} {\bibfnamefont {M.}~\bibnamefont {Arnaud}},
  \bibinfo {author} {\bibfnamefont {F.}~\bibnamefont {Arroja}}, \bibinfo
  {author} {\bibfnamefont {M.}~\bibnamefont {Ashdown}}, \bibinfo {author}
  {\bibfnamefont {J.}~\bibnamefont {Aumont}}, \bibinfo {author} {\bibfnamefont
  {C.}~\bibnamefont {Baccigalupi}}, \bibinfo {author} {\bibfnamefont
  {M.}~\bibnamefont {Ballardini}}, \bibinfo {author} {\bibfnamefont {A.~J.}\
  \bibnamefont {Banday}}, \ and\ \bibinfo {author} {\bibnamefont {et~al.}},\
  }\href {\doibase 10.1051/0004-6361/201525898} {\bibfield  {journal} {\bibinfo
   {journal} {Astronomy \& Astrophysics}\ }\textbf {\bibinfo {volume} {594}},\
  \bibinfo {pages} {A20} (\bibinfo {year} {2016})}\BibitemShut {NoStop}%
\bibitem [{\citenamefont {Martin}\ \emph {et~al.}(2014)\citenamefont {Martin},
  \citenamefont {Ringeval}, \citenamefont {Trotta},\ and\ \citenamefont
  {Vennin}}]{Martin:2013nzq}%
  \BibitemOpen
  \bibfield  {author} {\bibinfo {author} {\bibfnamefont {J.}~\bibnamefont
  {Martin}}, \bibinfo {author} {\bibfnamefont {C.}~\bibnamefont {Ringeval}},
  \bibinfo {author} {\bibfnamefont {R.}~\bibnamefont {Trotta}}, \ and\ \bibinfo
  {author} {\bibfnamefont {V.}~\bibnamefont {Vennin}},\ }\href {\doibase
  10.1088/1475-7516/2014/03/039} {\bibfield  {journal} {\bibinfo  {journal}
  {JCAP}\ }\textbf {\bibinfo {volume} {1403}},\ \bibinfo {pages} {039}
  (\bibinfo {year} {2014})},\ \Eprint {http://arxiv.org/abs/1312.3529}
  {arXiv:1312.3529 [astro-ph.CO]} \BibitemShut {NoStop}%
\bibitem [{\citenamefont {Brandenberger}\ and\ \citenamefont
  {Kung}(1990)}]{Brandenberger:1990wu}%
  \BibitemOpen
  \bibfield  {author} {\bibinfo {author} {\bibfnamefont {R.~H.}\ \bibnamefont
  {Brandenberger}}\ and\ \bibinfo {author} {\bibfnamefont {J.~H.}\ \bibnamefont
  {Kung}},\ }\href {\doibase 10.1103/PhysRevD.42.1008} {\bibfield  {journal}
  {\bibinfo  {journal} {Phys. Rev.}\ }\textbf {\bibinfo {volume} {D42}},\
  \bibinfo {pages} {1008} (\bibinfo {year} {1990})}\BibitemShut {NoStop}%
\bibitem [{\citenamefont {Brandenberger}\ \emph {et~al.}(1991)\citenamefont
  {Brandenberger}, \citenamefont {Feldman},\ and\ \citenamefont
  {Kung}}]{Brandenberger:1990xu}%
  \BibitemOpen
  \bibfield  {author} {\bibinfo {author} {\bibfnamefont {R.~H.}\ \bibnamefont
  {Brandenberger}}, \bibinfo {author} {\bibfnamefont {H.}~\bibnamefont
  {Feldman}}, \ and\ \bibinfo {author} {\bibfnamefont {J.}~\bibnamefont
  {Kung}},\ }\bibfield  {booktitle} {\emph {\bibinfo {booktitle} {{Nobel
  Symposium 1990: The Birth and Early Evolution of the Universe Graftavallen,
  Sweden, June 11-16, 1990}}},\ }\href {\doibase
  10.1088/0031-8949/1991/T36/007} {\bibfield  {journal} {\bibinfo  {journal}
  {Phys. Scripta}\ }\textbf {\bibinfo {volume} {T36}},\ \bibinfo {pages} {64}
  (\bibinfo {year} {1991})}\BibitemShut {NoStop}%
\bibitem [{\citenamefont {Alho}\ and\ \citenamefont {Mena}(2014)}]{Alho_2014}%
  \BibitemOpen
  \bibfield  {author} {\bibinfo {author} {\bibfnamefont {A.}~\bibnamefont
  {Alho}}\ and\ \bibinfo {author} {\bibfnamefont {F.~C.}\ \bibnamefont
  {Mena}},\ }\href {\doibase 10.1103/physrevd.90.043501} {\bibfield  {journal}
  {\bibinfo  {journal} {Physical Review D}\ }\textbf {\bibinfo {volume} {90}}
  (\bibinfo {year} {2014}),\ 10.1103/physrevd.90.043501}\BibitemShut {NoStop}%
\bibitem [{\citenamefont {Alho}\ and\ \citenamefont
  {Mena}(2011)}]{ALHO2011537}%
  \BibitemOpen
  \bibfield  {author} {\bibinfo {author} {\bibfnamefont {A.}~\bibnamefont
  {Alho}}\ and\ \bibinfo {author} {\bibfnamefont {F.~C.}\ \bibnamefont
  {Mena}},\ }\href {\doibase https://doi.org/10.1016/j.physletb.2011.08.044}
  {\bibfield  {journal} {\bibinfo  {journal} {Physics Letters B}\ }\textbf
  {\bibinfo {volume} {703}},\ \bibinfo {pages} {537 } (\bibinfo {year}
  {2011})}\BibitemShut {NoStop}%
\bibitem [{\citenamefont {Deruelle}\ and\ \citenamefont
  {Goldwirth}(1995)}]{Deruelle:1994pa}%
  \BibitemOpen
  \bibfield  {author} {\bibinfo {author} {\bibfnamefont {N.}~\bibnamefont
  {Deruelle}}\ and\ \bibinfo {author} {\bibfnamefont {D.~S.}\ \bibnamefont
  {Goldwirth}},\ }\href {\doibase 10.1103/PhysRevD.51.1563} {\bibfield
  {journal} {\bibinfo  {journal} {Phys. Rev.}\ }\textbf {\bibinfo {volume}
  {D51}},\ \bibinfo {pages} {1563} (\bibinfo {year} {1995})},\ \Eprint
  {http://arxiv.org/abs/gr-qc/9409056} {arXiv:gr-qc/9409056 [gr-qc]}
  \BibitemShut {NoStop}%
\bibitem [{\citenamefont {Azhar}\ and\ \citenamefont
  {Kaiser}(2018)}]{Azhar_2018}%
  \BibitemOpen
  \bibfield  {author} {\bibinfo {author} {\bibfnamefont {F.}~\bibnamefont
  {Azhar}}\ and\ \bibinfo {author} {\bibfnamefont {D.~I.}\ \bibnamefont
  {Kaiser}},\ }\href {\doibase 10.1103/physrevd.98.063515} {\bibfield
  {journal} {\bibinfo  {journal} {Physical Review D}\ }\textbf {\bibinfo
  {volume} {98}} (\bibinfo {year} {2018}),\
  10.1103/physrevd.98.063515}\BibitemShut {NoStop}%
\bibitem [{\citenamefont {Bloomfield}\ \emph {et~al.}(2019)\citenamefont
  {Bloomfield}, \citenamefont {Fitzpatrick}, \citenamefont {Hilbert},\ and\
  \citenamefont {Kaiser}}]{Bloomfield_2019}%
  \BibitemOpen
  \bibfield  {author} {\bibinfo {author} {\bibfnamefont {J.~K.}\ \bibnamefont
  {Bloomfield}}, \bibinfo {author} {\bibfnamefont {P.}~\bibnamefont
  {Fitzpatrick}}, \bibinfo {author} {\bibfnamefont {K.}~\bibnamefont
  {Hilbert}}, \ and\ \bibinfo {author} {\bibfnamefont {D.~I.}\ \bibnamefont
  {Kaiser}},\ }\href {\doibase 10.1103/physrevd.100.063512} {\bibfield
  {journal} {\bibinfo  {journal} {Physical Review D}\ }\textbf {\bibinfo
  {volume} {100}} (\bibinfo {year} {2019}),\
  10.1103/physrevd.100.063512}\BibitemShut {NoStop}%
\bibitem [{\citenamefont {Goldwirth}\ and\ \citenamefont
  {Piran}(1989)}]{Goldwirth:1989vz}%
  \BibitemOpen
  \bibfield  {author} {\bibinfo {author} {\bibfnamefont {D.~S.}\ \bibnamefont
  {Goldwirth}}\ and\ \bibinfo {author} {\bibfnamefont {T.}~\bibnamefont
  {Piran}},\ }\href {\doibase 10.1103/PhysRevD.40.3263} {\bibfield  {journal}
  {\bibinfo  {journal} {Phys. Rev.}\ }\textbf {\bibinfo {volume} {D40}},\
  \bibinfo {pages} {3263} (\bibinfo {year} {1989})}\BibitemShut {NoStop}%
\bibitem [{\citenamefont {Laguna}\ \emph {et~al.}(1991)\citenamefont {Laguna},
  \citenamefont {Kurki-Suonio},\ and\ \citenamefont {Matzner}}]{Laguna:1991zs}%
  \BibitemOpen
  \bibfield  {author} {\bibinfo {author} {\bibfnamefont {P.}~\bibnamefont
  {Laguna}}, \bibinfo {author} {\bibfnamefont {H.}~\bibnamefont
  {Kurki-Suonio}}, \ and\ \bibinfo {author} {\bibfnamefont {R.~A.}\
  \bibnamefont {Matzner}},\ }\href {\doibase 10.1103/PhysRevD.44.3077}
  {\bibfield  {journal} {\bibinfo  {journal} {Phys. Rev.}\ }\textbf {\bibinfo
  {volume} {D44}},\ \bibinfo {pages} {3077} (\bibinfo {year}
  {1991})}\BibitemShut {NoStop}%
\bibitem [{\citenamefont {Kurki-Suonio}\ \emph {et~al.}(1993)\citenamefont
  {Kurki-Suonio}, \citenamefont {Laguna},\ and\ \citenamefont
  {Matzner}}]{KurkiSuonio:1993fg}%
  \BibitemOpen
  \bibfield  {author} {\bibinfo {author} {\bibfnamefont {H.}~\bibnamefont
  {Kurki-Suonio}}, \bibinfo {author} {\bibfnamefont {P.}~\bibnamefont
  {Laguna}}, \ and\ \bibinfo {author} {\bibfnamefont {R.~A.}\ \bibnamefont
  {Matzner}},\ }\href {\doibase 10.1103/PhysRevD.48.3611} {\bibfield  {journal}
  {\bibinfo  {journal} {Phys. Rev.}\ }\textbf {\bibinfo {volume} {D48}},\
  \bibinfo {pages} {3611} (\bibinfo {year} {1993})},\ \Eprint
  {http://arxiv.org/abs/astro-ph/9306009} {arXiv:astro-ph/9306009 [astro-ph]}
  \BibitemShut {NoStop}%
\bibitem [{\citenamefont {East}\ \emph {et~al.}(2016)\citenamefont {East},
  \citenamefont {Kleban}, \citenamefont {Linde},\ and\ \citenamefont
  {Senatore}}]{East:2015ggf}%
  \BibitemOpen
  \bibfield  {author} {\bibinfo {author} {\bibfnamefont {W.~E.}\ \bibnamefont
  {East}}, \bibinfo {author} {\bibfnamefont {M.}~\bibnamefont {Kleban}},
  \bibinfo {author} {\bibfnamefont {A.}~\bibnamefont {Linde}}, \ and\ \bibinfo
  {author} {\bibfnamefont {L.}~\bibnamefont {Senatore}},\ }\href {\doibase
  10.1088/1475-7516/2016/09/010} {\bibfield  {journal} {\bibinfo  {journal}
  {Journal of Cosmology and Astroparticle Physics}\ }\textbf {\bibinfo {volume}
  {2016}},\ \bibinfo {pages} {010} (\bibinfo {year} {2016})}\BibitemShut
  {NoStop}%
\bibitem [{\citenamefont {Clough}\ \emph {et~al.}(2017)\citenamefont {Clough},
  \citenamefont {Lim}, \citenamefont {DiNunno}, \citenamefont {Fischler},
  \citenamefont {Flauger},\ and\ \citenamefont {Paban}}]{Clough:2016ymm}%
  \BibitemOpen
  \bibfield  {author} {\bibinfo {author} {\bibfnamefont {K.}~\bibnamefont
  {Clough}}, \bibinfo {author} {\bibfnamefont {E.~A.}\ \bibnamefont {Lim}},
  \bibinfo {author} {\bibfnamefont {B.~S.}\ \bibnamefont {DiNunno}}, \bibinfo
  {author} {\bibfnamefont {W.}~\bibnamefont {Fischler}}, \bibinfo {author}
  {\bibfnamefont {R.}~\bibnamefont {Flauger}}, \ and\ \bibinfo {author}
  {\bibfnamefont {S.}~\bibnamefont {Paban}},\ }\href {\doibase
  10.1088/1475-7516/2017/09/025} {\bibfield  {journal} {\bibinfo  {journal}
  {JCAP}\ }\textbf {\bibinfo {volume} {1709}},\ \bibinfo {pages} {025}
  (\bibinfo {year} {2017})},\ \Eprint {http://arxiv.org/abs/1608.04408}
  {arXiv:1608.04408 [hep-th]} \BibitemShut {NoStop}%
\bibitem [{\citenamefont {Aurrekoetxea}\ \emph {et~al.}(2020)\citenamefont
  {Aurrekoetxea}, \citenamefont {Clough}, \citenamefont {Flauger},\ and\
  \citenamefont {Lim}}]{Aurrekoetxea_2020}%
  \BibitemOpen
  \bibfield  {author} {\bibinfo {author} {\bibfnamefont {J.~C.}\ \bibnamefont
  {Aurrekoetxea}}, \bibinfo {author} {\bibfnamefont {K.}~\bibnamefont
  {Clough}}, \bibinfo {author} {\bibfnamefont {R.}~\bibnamefont {Flauger}}, \
  and\ \bibinfo {author} {\bibfnamefont {E.~A.}\ \bibnamefont {Lim}},\ }\href
  {\doibase 10.1088/1475-7516/2020/05/030} {\bibfield  {journal} {\bibinfo
  {journal} {Journal of Cosmology and Astroparticle Physics}\ }\textbf
  {\bibinfo {volume} {2020}},\ \bibinfo {pages} {030–030} (\bibinfo {year}
  {2020})}\BibitemShut {NoStop}%
\bibitem [{\citenamefont {Giblin}\ and\ \citenamefont
  {Tishue}(2019)}]{PhysRevD.100.063543}%
  \BibitemOpen
  \bibfield  {author} {\bibinfo {author} {\bibfnamefont {J.~T.}\ \bibnamefont
  {Giblin}}\ and\ \bibinfo {author} {\bibfnamefont {A.~J.}\ \bibnamefont
  {Tishue}},\ }\href {\doibase 10.1103/PhysRevD.100.063543} {\bibfield
  {journal} {\bibinfo  {journal} {Phys. Rev. D}\ }\textbf {\bibinfo {volume}
  {100}},\ \bibinfo {pages} {063543} (\bibinfo {year} {2019})}\BibitemShut
  {NoStop}%
\bibitem [{\citenamefont {Garfinkle}\ \emph {et~al.}(2008)\citenamefont
  {Garfinkle}, \citenamefont {Lim}, \citenamefont {Pretorius},\ and\
  \citenamefont {Steinhardt}}]{PhysRevD.78.083537}%
  \BibitemOpen
  \bibfield  {author} {\bibinfo {author} {\bibfnamefont {D.}~\bibnamefont
  {Garfinkle}}, \bibinfo {author} {\bibfnamefont {W.~C.}\ \bibnamefont {Lim}},
  \bibinfo {author} {\bibfnamefont {F.}~\bibnamefont {Pretorius}}, \ and\
  \bibinfo {author} {\bibfnamefont {P.~J.}\ \bibnamefont {Steinhardt}},\ }\href
  {\doibase 10.1103/PhysRevD.78.083537} {\bibfield  {journal} {\bibinfo
  {journal} {Phys. Rev. D}\ }\textbf {\bibinfo {volume} {78}},\ \bibinfo
  {pages} {083537} (\bibinfo {year} {2008})}\BibitemShut {NoStop}%
\bibitem [{\citenamefont {Ijjas}\ \emph {et~al.}(2020)\citenamefont {Ijjas},
  \citenamefont {Cook}, \citenamefont {Pretorius}, \citenamefont {Steinhardt},\
  and\ \citenamefont {Davies}}]{Ijjas_2020}%
  \BibitemOpen
  \bibfield  {author} {\bibinfo {author} {\bibfnamefont {A.}~\bibnamefont
  {Ijjas}}, \bibinfo {author} {\bibfnamefont {W.~G.}\ \bibnamefont {Cook}},
  \bibinfo {author} {\bibfnamefont {F.}~\bibnamefont {Pretorius}}, \bibinfo
  {author} {\bibfnamefont {P.~J.}\ \bibnamefont {Steinhardt}}, \ and\ \bibinfo
  {author} {\bibfnamefont {E.~Y.}\ \bibnamefont {Davies}},\ }\href {\doibase
  10.1088/1475-7516/2020/08/030} {\bibfield  {journal} {\bibinfo  {journal}
  {Journal of Cosmology and Astroparticle Physics}\ }\textbf {\bibinfo {volume}
  {2020}},\ \bibinfo {pages} {030} (\bibinfo {year} {2020})}\BibitemShut
  {NoStop}%
\bibitem [{\citenamefont {Goldwirth}(1991)}]{Goldwirth:1990pm}%
  \BibitemOpen
  \bibfield  {author} {\bibinfo {author} {\bibfnamefont {D.~S.}\ \bibnamefont
  {Goldwirth}},\ }\href {\doibase 10.1103/PhysRevD.43.3204} {\bibfield
  {journal} {\bibinfo  {journal} {Phys. Rev.}\ }\textbf {\bibinfo {volume}
  {D43}},\ \bibinfo {pages} {3204} (\bibinfo {year} {1991})}\BibitemShut
  {NoStop}%
\bibitem [{\citenamefont {Goldwirth}\ and\ \citenamefont
  {Piran}(1990)}]{Goldwirth:1989pr}%
  \BibitemOpen
  \bibfield  {author} {\bibinfo {author} {\bibfnamefont {D.~S.}\ \bibnamefont
  {Goldwirth}}\ and\ \bibinfo {author} {\bibfnamefont {T.}~\bibnamefont
  {Piran}},\ }\href {\doibase 10.1103/PhysRevLett.64.2852} {\bibfield
  {journal} {\bibinfo  {journal} {Phys. Rev. Lett.}\ }\textbf {\bibinfo
  {volume} {64}},\ \bibinfo {pages} {2852} (\bibinfo {year}
  {1990})}\BibitemShut {NoStop}%
\bibitem [{\citenamefont {Brandenberger}(2017)}]{Brandenberger:2016uzh}%
  \BibitemOpen
  \bibfield  {author} {\bibinfo {author} {\bibfnamefont {R.}~\bibnamefont
  {Brandenberger}},\ }\href {\doibase 10.1142/s0218271817400028} {\bibfield
  {journal} {\bibinfo  {journal} {International Journal of Modern Physics D}\
  }\textbf {\bibinfo {volume} {26}},\ \bibinfo {pages} {1740002} (\bibinfo
  {year} {2017})}\BibitemShut {NoStop}%
\bibitem [{\citenamefont {Clough}\ \emph {et~al.}(2015)\citenamefont {Clough},
  \citenamefont {Figueras}, \citenamefont {Finkel}, \citenamefont {Kunesch},
  \citenamefont {Lim},\ and\ \citenamefont {Tunyasuvunakool}}]{Clough_2015}%
  \BibitemOpen
  \bibfield  {author} {\bibinfo {author} {\bibfnamefont {K.}~\bibnamefont
  {Clough}}, \bibinfo {author} {\bibfnamefont {P.}~\bibnamefont {Figueras}},
  \bibinfo {author} {\bibfnamefont {H.}~\bibnamefont {Finkel}}, \bibinfo
  {author} {\bibfnamefont {M.}~\bibnamefont {Kunesch}}, \bibinfo {author}
  {\bibfnamefont {E.~A.}\ \bibnamefont {Lim}}, \ and\ \bibinfo {author}
  {\bibfnamefont {S.}~\bibnamefont {Tunyasuvunakool}},\ }\href {\doibase
  10.1088/0264-9381/32/24/245011} {\bibfield  {journal} {\bibinfo  {journal}
  {Classical and Quantum Gravity}\ }\textbf {\bibinfo {volume} {32}},\ \bibinfo
  {pages} {245011} (\bibinfo {year} {2015})},\ \Eprint
  {http://arxiv.org/abs/1503.03436} {1503.03436} \BibitemShut {NoStop}%
\bibitem [{\citenamefont {Shibata}\ and\ \citenamefont
  {Nakamura}(1995)}]{PhysRevD.52.5428}%
  \BibitemOpen
  \bibfield  {author} {\bibinfo {author} {\bibfnamefont {M.}~\bibnamefont
  {Shibata}}\ and\ \bibinfo {author} {\bibfnamefont {T.}~\bibnamefont
  {Nakamura}},\ }\href {\doibase 10.1103/PhysRevD.52.5428} {\bibfield
  {journal} {\bibinfo  {journal} {Phys. Rev. D}\ }\textbf {\bibinfo {volume}
  {52}},\ \bibinfo {pages} {5428} (\bibinfo {year} {1995})}\BibitemShut
  {NoStop}%
\bibitem [{\citenamefont {Baumgarte}\ and\ \citenamefont
  {Shapiro}(1998)}]{Baumgarte_1998}%
  \BibitemOpen
  \bibfield  {author} {\bibinfo {author} {\bibfnamefont {T.~W.}\ \bibnamefont
  {Baumgarte}}\ and\ \bibinfo {author} {\bibfnamefont {S.~L.}\ \bibnamefont
  {Shapiro}},\ }\href {\doibase 10.1103/physrevd.59.024007} {\bibfield
  {journal} {\bibinfo  {journal} {Physical Review D}\ }\textbf {\bibinfo
  {volume} {59}} (\bibinfo {year} {1998}),\
  10.1103/physrevd.59.024007}\BibitemShut {NoStop}%
\bibitem [{\citenamefont {Nakamura}\ \emph {et~al.}(1987)\citenamefont
  {Nakamura}, \citenamefont {Oohara},\ and\ \citenamefont
  {Kojima}}]{10.1143/PTPS.90.1}%
  \BibitemOpen
  \bibfield  {author} {\bibinfo {author} {\bibfnamefont {T.}~\bibnamefont
  {Nakamura}}, \bibinfo {author} {\bibfnamefont {K.}~\bibnamefont {Oohara}}, \
  and\ \bibinfo {author} {\bibfnamefont {Y.}~\bibnamefont {Kojima}},\ }\href
  {\doibase 10.1143/PTPS.90.1} {\bibfield  {journal} {\bibinfo  {journal}
  {Progress of Theoretical Physics Supplement}\ }\textbf {\bibinfo {volume}
  {90}},\ \bibinfo {pages} {1} (\bibinfo {year} {1987})}\BibitemShut {NoStop}%
\bibitem [{\citenamefont {Helfer}\ \emph {et~al.}(2019)\citenamefont {Helfer},
  \citenamefont {Aurrekoetxea},\ and\ \citenamefont {Lim}}]{Helfer_2019}%
  \BibitemOpen
  \bibfield  {author} {\bibinfo {author} {\bibfnamefont {T.}~\bibnamefont
  {Helfer}}, \bibinfo {author} {\bibfnamefont {J.~C.}\ \bibnamefont
  {Aurrekoetxea}}, \ and\ \bibinfo {author} {\bibfnamefont {E.~A.}\
  \bibnamefont {Lim}},\ }\href {\doibase 10.1103/physrevd.99.104028} {\bibfield
   {journal} {\bibinfo  {journal} {Physical Review D}\ }\textbf {\bibinfo
  {volume} {99}} (\bibinfo {year} {2019}),\ 10.1103/physrevd.99.104028},\
  \Eprint {http://arxiv.org/abs/1808.06678} {1808.06678} \BibitemShut {NoStop}%
\bibitem [{\citenamefont {Rekier}\ \emph {et~al.}(2015)\citenamefont {Rekier},
  \citenamefont {Cordero-Carrión},\ and\ \citenamefont
  {Fuzfa}}]{Rekier:2014rqa}%
  \BibitemOpen
  \bibfield  {author} {\bibinfo {author} {\bibfnamefont {J.}~\bibnamefont
  {Rekier}}, \bibinfo {author} {\bibfnamefont {I.}~\bibnamefont
  {Cordero-Carrión}}, \ and\ \bibinfo {author} {\bibfnamefont
  {A.}~\bibnamefont {Fuzfa}},\ }\href {\doibase 10.1103/PhysRevD.91.024025}
  {\bibfield  {journal} {\bibinfo  {journal} {Phys. Rev.}\ }\textbf {\bibinfo
  {volume} {D91}},\ \bibinfo {pages} {024025} (\bibinfo {year} {2015})},\
  \Eprint {http://arxiv.org/abs/1409.3476} {arXiv:1409.3476 [gr-qc]}
  \BibitemShut {NoStop}%
\bibitem [{\citenamefont {Rekier}\ \emph {et~al.}(2016)\citenamefont {Rekier},
  \citenamefont {Fuzfa},\ and\ \citenamefont
  {Cordero-Carrion}}]{Rekier:2015isa}%
  \BibitemOpen
  \bibfield  {author} {\bibinfo {author} {\bibfnamefont {J.}~\bibnamefont
  {Rekier}}, \bibinfo {author} {\bibfnamefont {A.}~\bibnamefont {Fuzfa}}, \
  and\ \bibinfo {author} {\bibfnamefont {I.}~\bibnamefont {Cordero-Carrion}},\
  }\href {\doibase 10.1103/PhysRevD.93.043533} {\bibfield  {journal} {\bibinfo
  {journal} {Phys. Rev.}\ }\textbf {\bibinfo {volume} {D93}},\ \bibinfo {pages}
  {043533} (\bibinfo {year} {2016})},\ \Eprint
  {http://arxiv.org/abs/1509.08354} {arXiv:1509.08354 [gr-qc]} \BibitemShut
  {NoStop}%
\bibitem [{\citenamefont {Linde}(2018)}]{Linde:2017pwt}%
  \BibitemOpen
  \bibfield  {author} {\bibinfo {author} {\bibfnamefont {A.}~\bibnamefont
  {Linde}},\ }\href {\doibase 10.1007/s10701-018-0177-9} {\bibfield  {journal}
  {\bibinfo  {journal} {Found. Phys.}\ }\textbf {\bibinfo {volume} {48}},\
  \bibinfo {pages} {1246} (\bibinfo {year} {2018})},\ \Eprint
  {http://arxiv.org/abs/1710.04278} {arXiv:1710.04278 [hep-th]} \BibitemShut
  {NoStop}%
\bibitem [{\citenamefont {Finn}\ and\ \citenamefont
  {Karamitsos}(2019)}]{Finn:2018krt}%
  \BibitemOpen
  \bibfield  {author} {\bibinfo {author} {\bibfnamefont {K.}~\bibnamefont
  {Finn}}\ and\ \bibinfo {author} {\bibfnamefont {S.}~\bibnamefont
  {Karamitsos}},\ }\href {\doibase 10.1103/PhysRevD.99.063515} {\bibfield
  {journal} {\bibinfo  {journal} {Phys. Rev. D}\ }\textbf {\bibinfo {volume}
  {99}},\ \bibinfo {pages} {063515} (\bibinfo {year} {2019})},\ \bibinfo {note}
  {[Erratum: Phys.Rev.D 99, 109901 (2019)]},\ \Eprint
  {http://arxiv.org/abs/1812.07095} {arXiv:1812.07095 [gr-qc]} \BibitemShut
  {NoStop}%
\bibitem [{\citenamefont {Chowdhury}\ \emph {et~al.}(2019)\citenamefont
  {Chowdhury}, \citenamefont {Martin}, \citenamefont {Ringeval},\ and\
  \citenamefont {Vennin}}]{Chowdhury:2019otk}%
  \BibitemOpen
  \bibfield  {author} {\bibinfo {author} {\bibfnamefont {D.}~\bibnamefont
  {Chowdhury}}, \bibinfo {author} {\bibfnamefont {J.}~\bibnamefont {Martin}},
  \bibinfo {author} {\bibfnamefont {C.}~\bibnamefont {Ringeval}}, \ and\
  \bibinfo {author} {\bibfnamefont {V.}~\bibnamefont {Vennin}},\ }\href
  {\doibase 10.1103/PhysRevD.100.083537} {\bibfield  {journal} {\bibinfo
  {journal} {Phys. Rev. D}\ }\textbf {\bibinfo {volume} {100}},\ \bibinfo
  {pages} {083537} (\bibinfo {year} {2019})},\ \Eprint
  {http://arxiv.org/abs/1902.03951} {arXiv:1902.03951 [astro-ph.CO]}
  \BibitemShut {NoStop}%
\bibitem [{\citenamefont {Tenkanen}\ and\ \citenamefont
  {Tomberg}(2020)}]{Tenkanen:2020cvw}%
  \BibitemOpen
  \bibfield  {author} {\bibinfo {author} {\bibfnamefont {T.}~\bibnamefont
  {Tenkanen}}\ and\ \bibinfo {author} {\bibfnamefont {E.}~\bibnamefont
  {Tomberg}},\ }\href {\doibase 10.1088/1475-7516/2020/04/050} {\bibfield
  {journal} {\bibinfo  {journal} {JCAP}\ }\textbf {\bibinfo {volume} {04}},\
  \bibinfo {pages} {050} (\bibinfo {year} {2020})},\ \Eprint
  {http://arxiv.org/abs/2002.02420} {arXiv:2002.02420 [astro-ph.CO]}
  \BibitemShut {NoStop}%
\bibitem [{\citenamefont {Goldwirth}\ and\ \citenamefont
  {Piran}(1992)}]{Goldwirth:1991rj}%
  \BibitemOpen
  \bibfield  {author} {\bibinfo {author} {\bibfnamefont {D.~S.}\ \bibnamefont
  {Goldwirth}}\ and\ \bibinfo {author} {\bibfnamefont {T.}~\bibnamefont
  {Piran}},\ }\href {\doibase 10.1016/0370-1573(92)90073-9} {\bibfield
  {journal} {\bibinfo  {journal} {Phys. Rept.}\ }\textbf {\bibinfo {volume}
  {214}},\ \bibinfo {pages} {223} (\bibinfo {year} {1992})}\BibitemShut
  {NoStop}%
\bibitem [{\citenamefont {York}(1971)}]{York:1971hw}%
  \BibitemOpen
  \bibfield  {author} {\bibinfo {author} {\bibfnamefont {J.}~\bibnamefont
  {York}, \bibfnamefont {James~W.}},\ }\href {\doibase
  10.1103/PhysRevLett.26.1656} {\bibfield  {journal} {\bibinfo  {journal}
  {Phys. Rev. Lett.}\ }\textbf {\bibinfo {volume} {26}},\ \bibinfo {pages}
  {1656} (\bibinfo {year} {1971})}\BibitemShut {NoStop}%
\bibitem [{\citenamefont {York}(1972)}]{York:1972sj}%
  \BibitemOpen
  \bibfield  {author} {\bibinfo {author} {\bibfnamefont {J.}~\bibnamefont
  {York}, \bibfnamefont {James~W.}},\ }\href {\doibase
  10.1103/PhysRevLett.28.1082} {\bibfield  {journal} {\bibinfo  {journal}
  {Phys. Rev. Lett.}\ }\textbf {\bibinfo {volume} {28}},\ \bibinfo {pages}
  {1082} (\bibinfo {year} {1972})}\BibitemShut {NoStop}%
\bibitem [{\citenamefont {Ijjas}\ \emph {et~al.}(2013)\citenamefont {Ijjas},
  \citenamefont {Steinhardt},\ and\ \citenamefont {Loeb}}]{Ijjas_2013}%
  \BibitemOpen
  \bibfield  {author} {\bibinfo {author} {\bibfnamefont {A.}~\bibnamefont
  {Ijjas}}, \bibinfo {author} {\bibfnamefont {P.~J.}\ \bibnamefont
  {Steinhardt}}, \ and\ \bibinfo {author} {\bibfnamefont {A.}~\bibnamefont
  {Loeb}},\ }\href {\doibase 10.1016/j.physletb.2013.05.023} {\bibfield
  {journal} {\bibinfo  {journal} {Physics Letters B}\ }\textbf {\bibinfo
  {volume} {723}},\ \bibinfo {pages} {261–266} (\bibinfo {year}
  {2013})}\BibitemShut {NoStop}%
\bibitem [{\citenamefont {Guth}\ \emph {et~al.}(2014)\citenamefont {Guth},
  \citenamefont {Kaiser},\ and\ \citenamefont {Nomura}}]{Guth_2014}%
  \BibitemOpen
  \bibfield  {author} {\bibinfo {author} {\bibfnamefont {A.~H.}\ \bibnamefont
  {Guth}}, \bibinfo {author} {\bibfnamefont {D.~I.}\ \bibnamefont {Kaiser}}, \
  and\ \bibinfo {author} {\bibfnamefont {Y.}~\bibnamefont {Nomura}},\ }\href
  {\doibase 10.1016/j.physletb.2014.03.020} {\bibfield  {journal} {\bibinfo
  {journal} {Physics Letters B}\ }\textbf {\bibinfo {volume} {733}},\ \bibinfo
  {pages} {112–119} (\bibinfo {year} {2014})}\BibitemShut {NoStop}%
\bibitem [{\citenamefont {Ijjas}\ and\ \citenamefont
  {Steinhardt}(2016)}]{Ijjas_2016}%
  \BibitemOpen
  \bibfield  {author} {\bibinfo {author} {\bibfnamefont {A.}~\bibnamefont
  {Ijjas}}\ and\ \bibinfo {author} {\bibfnamefont {P.~J.}\ \bibnamefont
  {Steinhardt}},\ }\href {\doibase 10.1088/0264-9381/33/4/044001} {\bibfield
  {journal} {\bibinfo  {journal} {Classical and Quantum Gravity}\ }\textbf
  {\bibinfo {volume} {33}},\ \bibinfo {pages} {044001} (\bibinfo {year}
  {2016})}\BibitemShut {NoStop}%
\bibitem [{\citenamefont {Easther}\ \emph {et~al.}(2014)\citenamefont
  {Easther}, \citenamefont {Price},\ and\ \citenamefont
  {Rasero}}]{Easther:2014zga}%
  \BibitemOpen
  \bibfield  {author} {\bibinfo {author} {\bibfnamefont {R.}~\bibnamefont
  {Easther}}, \bibinfo {author} {\bibfnamefont {L.~C.}\ \bibnamefont {Price}},
  \ and\ \bibinfo {author} {\bibfnamefont {J.}~\bibnamefont {Rasero}},\ }\href
  {\doibase 10.1088/1475-7516/2014/08/041} {\bibfield  {journal} {\bibinfo
  {journal} {JCAP}\ }\textbf {\bibinfo {volume} {1408}},\ \bibinfo {pages}
  {041} (\bibinfo {year} {2014})},\ \Eprint {http://arxiv.org/abs/1406.2869}
  {arXiv:1406.2869 [astro-ph.CO]} \BibitemShut {NoStop}%
\bibitem [{\citenamefont {Clesse}\ \emph {et~al.}(2009)\citenamefont {Clesse},
  \citenamefont {Ringeval},\ and\ \citenamefont {Rocher}}]{Clesse:2009ur}%
  \BibitemOpen
  \bibfield  {author} {\bibinfo {author} {\bibfnamefont {S.}~\bibnamefont
  {Clesse}}, \bibinfo {author} {\bibfnamefont {C.}~\bibnamefont {Ringeval}}, \
  and\ \bibinfo {author} {\bibfnamefont {J.}~\bibnamefont {Rocher}},\ }\href
  {\doibase 10.1103/PhysRevD.80.123534} {\bibfield  {journal} {\bibinfo
  {journal} {Phys. Rev.}\ }\textbf {\bibinfo {volume} {D80}},\ \bibinfo {pages}
  {123534} (\bibinfo {year} {2009})},\ \Eprint {http://arxiv.org/abs/0909.0402}
  {arXiv:0909.0402 [astro-ph.CO]} \BibitemShut {NoStop}%
\bibitem [{\citenamefont {Clesse}\ and\ \citenamefont
  {Rocher}(2009)}]{Clesse:2008pf}%
  \BibitemOpen
  \bibfield  {author} {\bibinfo {author} {\bibfnamefont {S.}~\bibnamefont
  {Clesse}}\ and\ \bibinfo {author} {\bibfnamefont {J.}~\bibnamefont
  {Rocher}},\ }\href {\doibase 10.1103/PhysRevD.79.103507} {\bibfield
  {journal} {\bibinfo  {journal} {Phys. Rev.}\ }\textbf {\bibinfo {volume}
  {D79}},\ \bibinfo {pages} {103507} (\bibinfo {year} {2009})},\ \Eprint
  {http://arxiv.org/abs/0809.4355} {arXiv:0809.4355 [hep-ph]} \BibitemShut
  {NoStop}%
\bibitem [{\citenamefont {Clough}\ \emph {et~al.}(2018)\citenamefont {Clough},
  \citenamefont {Flauger},\ and\ \citenamefont {Lim}}]{Clough_2018}%
  \BibitemOpen
  \bibfield  {author} {\bibinfo {author} {\bibfnamefont {K.}~\bibnamefont
  {Clough}}, \bibinfo {author} {\bibfnamefont {R.}~\bibnamefont {Flauger}}, \
  and\ \bibinfo {author} {\bibfnamefont {E.~A.}\ \bibnamefont {Lim}},\
  }\href@noop {} {\bibfield  {journal} {\bibinfo  {journal} {JCAP}\ }\textbf
  {\bibinfo {volume} {2018}},\ \bibinfo {pages} {065} (\bibinfo {year}
  {2018})},\ \Eprint {http://arxiv.org/abs/1712.07352} {1712.07352}
  \BibitemShut {NoStop}%
\bibitem [{\citenamefont {Bezrukov}\ and\ \citenamefont
  {Shaposhnikov}(2008)}]{Bezrukov:2007ep}%
  \BibitemOpen
  \bibfield  {author} {\bibinfo {author} {\bibfnamefont {F.~L.}\ \bibnamefont
  {Bezrukov}}\ and\ \bibinfo {author} {\bibfnamefont {M.}~\bibnamefont
  {Shaposhnikov}},\ }\href {\doibase 10.1016/j.physletb.2007.11.072} {\bibfield
   {journal} {\bibinfo  {journal} {Phys. Lett. B}\ }\textbf {\bibinfo {volume}
  {659}},\ \bibinfo {pages} {703} (\bibinfo {year} {2008})},\ \Eprint
  {http://arxiv.org/abs/0710.3755} {arXiv:0710.3755 [hep-th]} \BibitemShut
  {NoStop}%
\bibitem [{\citenamefont {Baker}\ \emph {et~al.}(2006)\citenamefont {Baker},
  \citenamefont {Centrella}, \citenamefont {Choi}, \citenamefont {Koppitz},\
  and\ \citenamefont {van Meter}}]{Baker_2006}%
  \BibitemOpen
  \bibfield  {author} {\bibinfo {author} {\bibfnamefont {J.~G.}\ \bibnamefont
  {Baker}}, \bibinfo {author} {\bibfnamefont {J.}~\bibnamefont {Centrella}},
  \bibinfo {author} {\bibfnamefont {D.-I.}\ \bibnamefont {Choi}}, \bibinfo
  {author} {\bibfnamefont {M.}~\bibnamefont {Koppitz}}, \ and\ \bibinfo
  {author} {\bibfnamefont {J.}~\bibnamefont {van Meter}},\ }\href {\doibase
  10.1103/physrevlett.96.111102} {\bibfield  {journal} {\bibinfo  {journal}
  {Physical Review Letters}\ }\textbf {\bibinfo {volume} {96}} (\bibinfo {year}
  {2006}),\ 10.1103/physrevlett.96.111102}\BibitemShut {NoStop}%
\bibitem [{\citenamefont {Campanelli}\ \emph {et~al.}(2006)\citenamefont
  {Campanelli}, \citenamefont {Lousto}, \citenamefont {Marronetti},\ and\
  \citenamefont {Zlochower}}]{Campanelli_2006}%
  \BibitemOpen
  \bibfield  {author} {\bibinfo {author} {\bibfnamefont {M.}~\bibnamefont
  {Campanelli}}, \bibinfo {author} {\bibfnamefont {C.~O.}\ \bibnamefont
  {Lousto}}, \bibinfo {author} {\bibfnamefont {P.}~\bibnamefont {Marronetti}},
  \ and\ \bibinfo {author} {\bibfnamefont {Y.}~\bibnamefont {Zlochower}},\
  }\href {\doibase 10.1103/physrevlett.96.111101} {\bibfield  {journal}
  {\bibinfo  {journal} {Physical Review Letters}\ }\textbf {\bibinfo {volume}
  {96}} (\bibinfo {year} {2006}),\ 10.1103/physrevlett.96.111101}\BibitemShut
  {NoStop}%
\bibitem [{\citenamefont {Garfinkle}\ and\ \citenamefont
  {Mead}(2020)}]{Garfinkle_2020}%
  \BibitemOpen
  \bibfield  {author} {\bibinfo {author} {\bibfnamefont {D.}~\bibnamefont
  {Garfinkle}}\ and\ \bibinfo {author} {\bibfnamefont {L.}~\bibnamefont
  {Mead}},\ }\href {\doibase 10.1103/physrevd.102.044022} {\bibfield  {journal}
  {\bibinfo  {journal} {Physical Review D}\ }\textbf {\bibinfo {volume} {102}}
  (\bibinfo {year} {2020}),\ 10.1103/physrevd.102.044022}\BibitemShut {NoStop}%
\bibitem [{\citenamefont {Linde}(2004)}]{Linde:2004nz}%
  \BibitemOpen
  \bibfield  {author} {\bibinfo {author} {\bibfnamefont {A.~D.}\ \bibnamefont
  {Linde}},\ }\href {\doibase 10.1088/1475-7516/2004/10/004} {\bibfield
  {journal} {\bibinfo  {journal} {JCAP}\ }\textbf {\bibinfo {volume} {10}},\
  \bibinfo {pages} {004} (\bibinfo {year} {2004})},\ \Eprint
  {http://arxiv.org/abs/hep-th/0408164} {arXiv:hep-th/0408164} \BibitemShut
  {NoStop}%
\bibitem [{\citenamefont {Linde}(1983)}]{Linde:1983gd}%
  \BibitemOpen
  \bibfield  {author} {\bibinfo {author} {\bibfnamefont {A.~D.}\ \bibnamefont
  {Linde}},\ }\href {\doibase 10.1016/0370-2693(83)90837-7} {\bibfield
  {journal} {\bibinfo  {journal} {Phys. Lett. B}\ }\textbf {\bibinfo {volume}
  {129}},\ \bibinfo {pages} {177} (\bibinfo {year} {1983})}\BibitemShut
  {NoStop}%
\bibitem [{\citenamefont {Goncharov}\ and\ \citenamefont
  {Linde}(1984)}]{Goncharov:1983mw}%
  \BibitemOpen
  \bibfield  {author} {\bibinfo {author} {\bibfnamefont {A.}~\bibnamefont
  {Goncharov}}\ and\ \bibinfo {author} {\bibfnamefont {A.~D.}\ \bibnamefont
  {Linde}},\ }\href {\doibase 10.1016/0370-2693(84)90027-3} {\bibfield
  {journal} {\bibinfo  {journal} {Phys. Lett. B}\ }\textbf {\bibinfo {volume}
  {139}},\ \bibinfo {pages} {27} (\bibinfo {year} {1984})}\BibitemShut
  {NoStop}%
\bibitem [{\citenamefont {Linde}(1984)}]{Linde:1983mx}%
  \BibitemOpen
  \bibfield  {author} {\bibinfo {author} {\bibfnamefont {A.~D.}\ \bibnamefont
  {Linde}},\ }\href {\doibase 10.1007/BF02790571} {\bibfield  {journal}
  {\bibinfo  {journal} {Lett. Nuovo Cim.}\ }\textbf {\bibinfo {volume} {39}},\
  \bibinfo {pages} {401} (\bibinfo {year} {1984})}\BibitemShut {NoStop}%
\bibitem [{\citenamefont {Kofman}\ \emph {et~al.}(1985)\citenamefont {Kofman},
  \citenamefont {Linde},\ and\ \citenamefont {Starobinsky}}]{Kofman:1985aw}%
  \BibitemOpen
  \bibfield  {author} {\bibinfo {author} {\bibfnamefont {L.}~\bibnamefont
  {Kofman}}, \bibinfo {author} {\bibfnamefont {A.~D.}\ \bibnamefont {Linde}}, \
  and\ \bibinfo {author} {\bibfnamefont {A.~A.}\ \bibnamefont {Starobinsky}},\
  }\href {\doibase 10.1016/0370-2693(85)90381-8} {\bibfield  {journal}
  {\bibinfo  {journal} {Phys. Lett. B}\ }\textbf {\bibinfo {volume} {157}},\
  \bibinfo {pages} {361} (\bibinfo {year} {1985})}\BibitemShut {NoStop}%
\bibitem [{\citenamefont {Linde}(1987)}]{Linde:1987aa}%
  \BibitemOpen
  \bibfield  {author} {\bibinfo {author} {\bibfnamefont {A.~D.}\ \bibnamefont
  {Linde}},\ }\href {\doibase 10.1063/1.881088} {\bibfield  {journal} {\bibinfo
   {journal} {Phys. Today}\ }\textbf {\bibinfo {volume} {40}},\ \bibinfo
  {pages} {61} (\bibinfo {year} {1987})}\BibitemShut {NoStop}%
\bibitem [{\citenamefont {Bethel}\ \emph {et~al.}(2012)\citenamefont {Bethel},
  \citenamefont {Childs},\ and\ \citenamefont {Hansen}}]{10.5555/2422936}%
  \BibitemOpen
  \bibfield  {author} {\bibinfo {author} {\bibfnamefont {E.~W.}\ \bibnamefont
  {Bethel}}, \bibinfo {author} {\bibfnamefont {H.}~\bibnamefont {Childs}}, \
  and\ \bibinfo {author} {\bibfnamefont {C.}~\bibnamefont {Hansen}},\
  }\href@noop {} {\emph {\bibinfo {title} {High Performance Visualization:
  Enabling Extreme-Scale Scientific Insight}}},\ \bibinfo {edition} {1st}\ ed.\
  (\bibinfo  {publisher} {Chapman; Hall/CRC},\ \bibinfo {year}
  {2012})\BibitemShut {NoStop}%
\bibitem [{\citenamefont {Turk}\ \emph {et~al.}(2010)\citenamefont {Turk},
  \citenamefont {Smith}, \citenamefont {Oishi}, \citenamefont {Skory},
  \citenamefont {Skillman}, \citenamefont {Abel},\ and\ \citenamefont
  {Norman}}]{Turk_2010}%
  \BibitemOpen
  \bibfield  {author} {\bibinfo {author} {\bibfnamefont {M.~J.}\ \bibnamefont
  {Turk}}, \bibinfo {author} {\bibfnamefont {B.~D.}\ \bibnamefont {Smith}},
  \bibinfo {author} {\bibfnamefont {J.~S.}\ \bibnamefont {Oishi}}, \bibinfo
  {author} {\bibfnamefont {S.}~\bibnamefont {Skory}}, \bibinfo {author}
  {\bibfnamefont {S.~W.}\ \bibnamefont {Skillman}}, \bibinfo {author}
  {\bibfnamefont {T.}~\bibnamefont {Abel}}, \ and\ \bibinfo {author}
  {\bibfnamefont {M.~L.}\ \bibnamefont {Norman}},\ }\href {\doibase
  10.1088/0067-0049/192/1/9} {\bibfield  {journal} {\bibinfo  {journal} {The
  Astrophysical Journal Supplement Series}\ }\textbf {\bibinfo {volume}
  {192}},\ \bibinfo {pages} {9} (\bibinfo {year} {2010})}\BibitemShut {NoStop}%
\end{thebibliography}%
\end{document}